\documentclass[manuscript]{aastex}

\usepackage{amsmath}

\usepackage{color}

\shorttitle{Modelling Separator Reconnection}
\shortauthors{Wilmot-Smith \& Hornig}

\title{A time-dependent model for magnetic reconnection \\ in the  presence of a separator.}

\author{A.L. Wilmot-Smith }
\affil{Division of Mathematics, University of Dundee, 
Dundee, DD1 4HN, United Kingdom}
\email{antonia@maths.dundee.ac.uk}

\author{G. Hornig}
\affil{Division of Mathematics, University of Dundee, 
Dundee, DD1 4HN, United Kingdom}

\begin{document}

\begin{abstract}
We present a  model for separator reconnection due to an isolated reconnection process.  Separator reconnection is a process which occurs in the neighbourhood of a distinguished field line (the separator) connecting two null points of a magnetic field.  It is, for example, important for the dynamics of magnetic flux at the dayside magnetopause and in the solar corona.
We find that, above  a certain threshold, such a reconnection process generates new
separators which leads to a complex system of magnetic flux tubes connecting regions of previously separated flux.  Our findings are consistent with the findings of large numbers of separators in numerical simulations. We discuss how to measure and interpret the reconnection rate in a configuration with multiple separators.
\end{abstract}


\section{Introduction}

The magnetic field evolution in plasmas such as the solar corona and the Earth's Magnetosphere 
often follows  an ideal evolution in which all topological features of the field remain unchanged over time. This invariance provides a motivation to investigate topological properties of the magnetic field such as magnetic null points, flux surfaces and periodic field lines. 
Some of these topological features, particularly magnetic null points, are surprisingly stable and turn out to be conserved not only under an ideal plasma dynamics but also for a wide range of non-ideal flows. Mathematically this is reflected in the existence of an index theorem (Greene, 1992) for magnetic null points which allows only a small number of bifurcation processes to either generate or destroy magnetic null points. This underlines the importance of magnetic null points and their properties for the structure and dynamics of  astrophysical plasmas.

\begin{figure}[htbp] 
 \centering
\includegraphics[width=0.3\textwidth]{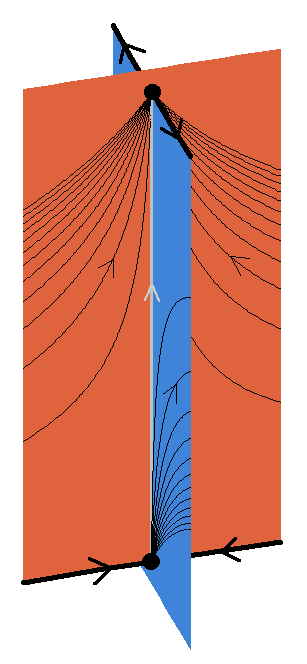} 
\caption{Two magnetic null points (black spheres) in a potential magnetic field configuration.
The fan planes of the upper null (orange) and lower null (blue) intersect along the separator (grey line).
Some particular  illustrative field lines  (thin black lines) in the fan planes are also shown, as 
are the spine lines of the nulls (thick black lines).}
\label{fig:initialstate2}
\end{figure}

Generic magnetic null points (i.e.~null points where all eigenvalues of the linearisation of the
magnetic field at the null have non-zero real part) are associated with certain distinguished field lines and flux surfaces, namely  
spines (or $\gamma$ lines in the terminology of Lau \& Finn, 1990) and fan surfaces (or $\Sigma$ surfaces)
as well as, sometimes, magnetic separators, the intersection lines of two fan surfaces.  
Combined, these features are sometimes termed the magnetic skeleton (e.g.~Priest \& Titov, 1996).
In dynamical systems terminology, null points are hyperbolic fixed points of a volume preserving (divergence-free) 
vector field and the spine and fan of a null are stable or unstable invariant manifolds.
A null whose fan surface corresponds to a  stable manifold and the spine line to an unstable manifold is designated as
Type~A while the converse is known as Type~B.  
The magnetic separator is a heteroclinic intersection of fan surfaces, lying in the fan plane of both a Type~A and Type~B null  as shown in Figure~\ref{fig:initialstate2}.

Conservation of topology has important implications for the plasma evolution, limiting the evolutions that may occur
and the amount of free magnetic energy bound in the configuration.
However, even for plasmas which are largely ideal, the frozen-in condition can be violated in localised regions
(such as current sheets) where non-ideal effects become important.   The key requirement 
for such a process to change the topology is to have a non-zero
electric field component in the direction parallel to the magnetic field (Hesse \& Schindler, 1988).
Such a component (for further details see Sections~4 and 5) enables a change in connectivity of the 
magnetic field lines and so a change of topology, the process being known as  magnetic reconnection.
Reconnection may  allow for significant magnetic energy release, both locally at the reconnection site and 
globally in the previously topologically bound energy.

Historically, models of magnetic reconnection were solely two-dimensional. Theoretical considerations show that
in two-dimensions reconnection can only occur at a hyperbolic (X-type) null-point of the field (for details see e.g.~Hornig, 2007).
Such null-points, lying at the intersection of four topologically distinct flux domains, are also likely
sites of current sheet formation.  The resultant reconnection  is now fairly well understood 
(for reviews see, e.g.~Biskamp  2000, Priest \& Forbes 2000), although modelling still required to fully determine 
the importance of the complete physics including kinetic effects (Birn \& Priest 2007).  

In three-dimensions, the typical situation in astrophysical plasmas,
 there are no fundamental restrictions on where reconnection occurs, the only requirement being
the presence of a localised non-ideal term in Ohm's law (Hesse \& Schindler, 2007; Schindler {\it et al.}~1988).
Accordingly, reconnection may take place wherever such non-ideal terms become important.  
Three-dimensional null-points are thought to be one such location 
(Klapper {\it et al.}~1996).  Various forms of reconnection can occur at 3D nulls (for a review see Priest \& Pontin 2009) 
and the nature of the reconnection has been analysed in some detail 
(e.g.~Pontin {\it et al.}~2004, 2005; Galsgaard \& Pontin 2011).

Current sheets may also form away from null points (e.g.~Titov {\it et al.}~2002;  Browning {\it et al.}~2008;  
Wilmot-Smith {\it et al.}~2009).
Models for the local reconnection process here (e.g.~Hornig \& Priest 2003, Wilmot-Smith {\it et al.}~2006, 2009) 
show that, just as in the 3D null case, there are many distinct 3D characteristics of reconnection, some of
which we discuss below.  One topologically distinguished location where current sheets may form away from  
nulls is at separators which, like the 2D null point, lie at the intersection of four flux domains.
It is this characteristic that has been used to argue why separators should be prone to current sheet formation
(Lau \& Finn 1990).  The minimum energy state for certain field configurations has the current lying along separators 
(Longcope 1996, 2001) and current sheet formation at separators has been observed in numerical simulations  
(e.g.~Galsgaard \& Nordlund, 1997; Haynes {\it et al.}~2007).   Reconnection taking place at separators is known as separator reconnection.

A recent series of papers (including Galsgaard \& Parnell 2005; Haynes {\it et al.}~2007; Parnell {\it et al.}~2008; Parnell {\it et al.}~2010a)
examine the reconnection taking place in a `fly-by' experiment where two magnetic flux patches on the photosphere are moved past each other and the resultant evolution in the corona is followed.
The authors find strong current concentrations and therefore reconnection around each of a number of separators 
(up to five in some time frames, Haynes {\it et al.}~2007).  
These separators are created in the reconnection process and connect the same two nulls 
(see also Longcope \& Cowley, 1996).  Magnetic flux is found to evolve in a complex manner, 
being reconnected multiple times through these distinct reconnection sites (Parnell {\it et al.}~2008).

Separators were also found to be important in a simulation of magnetic-flux emergence into the corona
(Parnell {\it et al.}~2010b) where current sheets between the emerging and pre-existing coronal field were shown 
to be threaded by numerous separators, a feature of complex magnetic fields suggested by Albright (1999).
Parnell {\it et al.}~(2010b) found the regions of highest integrated parallel electric field to be at or
in association with the separators, so arguing
that reconnection occurs at the separators themselves.
Separator reconnection also appears to be a key process at the magnetopause 
where magnetic nulls or even clusters of nulls appear in the northern and southern polar cusp regions
(Dorelli {\it et al.}~2007).  These nulls are linked by a separator along the dayside magnetopause
which will therefore be involved in reconnection with the interplanetary magnetic field.
Indeed reconnection in such a configuration has been implied by observations from the Cluster
spacecraft (Xiao {\it et al.}~2007).
These studies partially motivate  the present work where we want to investigate how and why multiple separators
 can be involved in reconnection.

Although separator configurations have been analysed for some time
(e.g.~Chance {\it et al.}~1992; Craig {\it et al.}~1999),
there remain several unanswered basic questions surrounding the nature of separator reconnection, 
particularly involving the way magnetic flux evolves, i.e.~exactly how the changes in field line connectivity occur.
There have been  suggestions that the reconnection involves a `cut and paste' of field lines at the separator itself 
in a manner similar to the 2D null case (e.g. Lau \& Finn 1990;  Priest \& Titov 1996; Longcope {\it et al.}~2001).
However, previous detailed investigations into the nature of individual 3D reconnection events both at and away 
from null points have shown the flux evolution to be quite distinct from the 2D picture 
(Hornig \& Priest 2003; Pontin {\it et al.}~2004, 2005) and, in the light of these models, such a simple flux evolution would 
be surprising.

Indeed a number of key features of 3D reconnection are now known which show many of its features
are quite distinct to the 2D case.  Accordingly it would appear useful to re-examine the fundamental nature of separator reconnection and consider whether it has its own particular distinguishing characteristics.
In doing so we also wish to address a conclusion of Parnell et al.~(2010a) who suggested that separator reconnection 
does not appear to involve the nulls that lie at both ends of the separator
as well as the recent findings (Parnell {\it et al.}~2010b) of very large numbers of separators sometimes appearing in 
numerical MHD experiments.

Accordingly the aim of the present work is to describe in detail the nature of an isolated 3D reconnection event
in the vicinity of a separator.  We do so using a simple analytical model which is described in detail in 
Section~\ref{ref:themodel}.
The model allows us to consider typical magnetic field connectivities resulting from separator reconnection 
in Section~\ref{sec:bif} and the nature of the magnetic flux evolution in Section~\ref{sec:evolve}.
We consider how reconnection rates may be determined in separator configurations
in Section~\ref{sec:rates} before discussing our findings and concluding in Sections~\ref{sec:discussion} \& \ref{sec:conc}.

\section{Introducing the model}
\label{ref:themodel}

A fully self-consistent model for  reconnection must incorporate a dynamic evolution 
which generates current sheet(s) as well as the reconnection that takes place at those current sheets 
and changes the magnetic field topology.  
An example in the solar corona is the emergence of a magnetic flux tube 
from the convection zone and reconnection with the pre-existing coronal magnetic field.   
Inherent in such events is an enormous separation of scales between the global dynamic process
and the local reconnection events.  Accordingly, a typical approach to model  reconnection 
itself is to start with a local magnetic field configuration that is considered susceptible to current sheet formation 
(such as, in two dimensions, an X-type null point of the field).  
In simulations the magnetic field is then confined to a finite region 
and the boundaries driven in such a manner as to initiate a reconnection event which can then be studied in detail.
Determining physically realistic boundary conditions is just one of the obstacles in this modelling technique.

\begin{figure}[htbp] 
   \centering
      \includegraphics[width=0.24\textwidth]{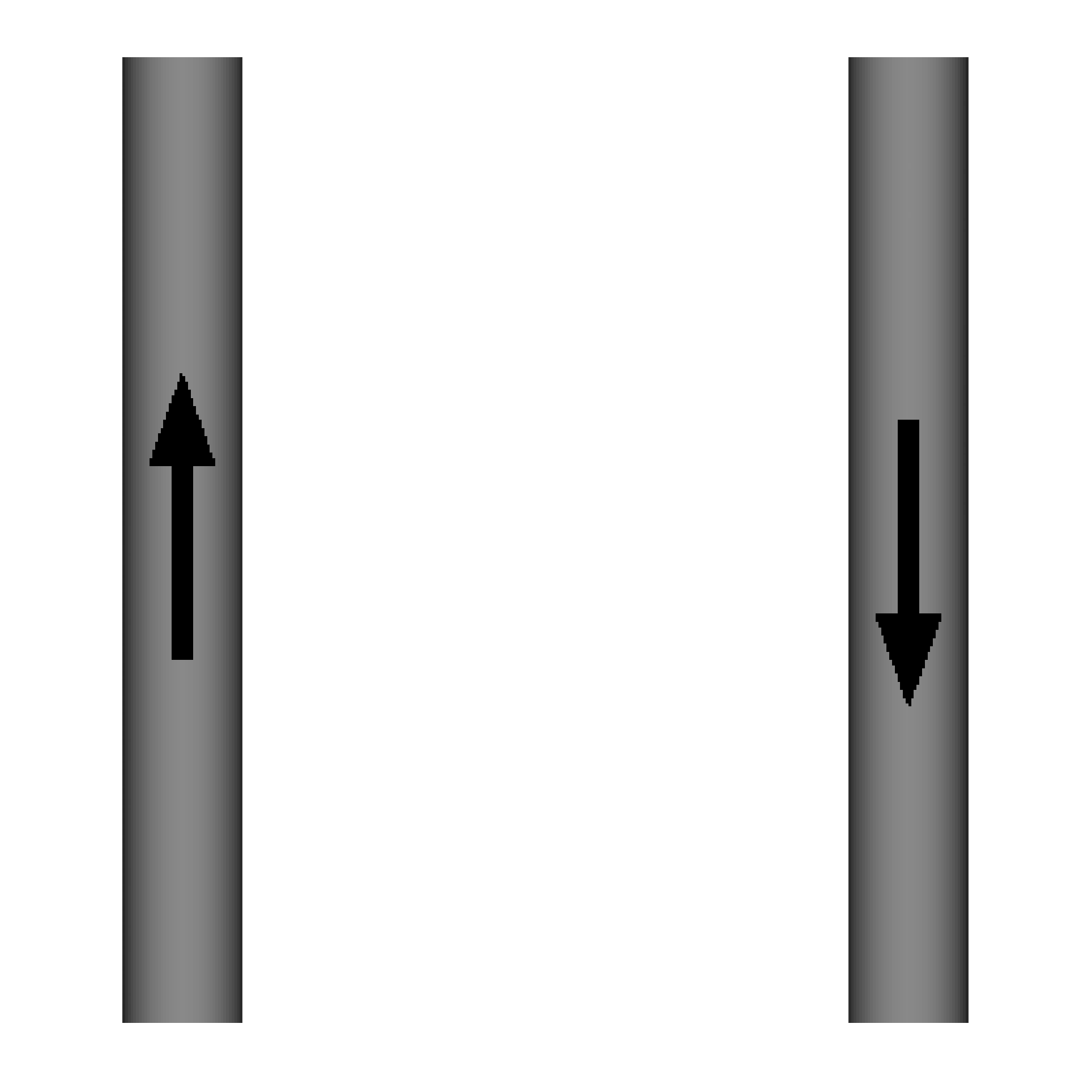}  
         \includegraphics[width=0.24\textwidth]{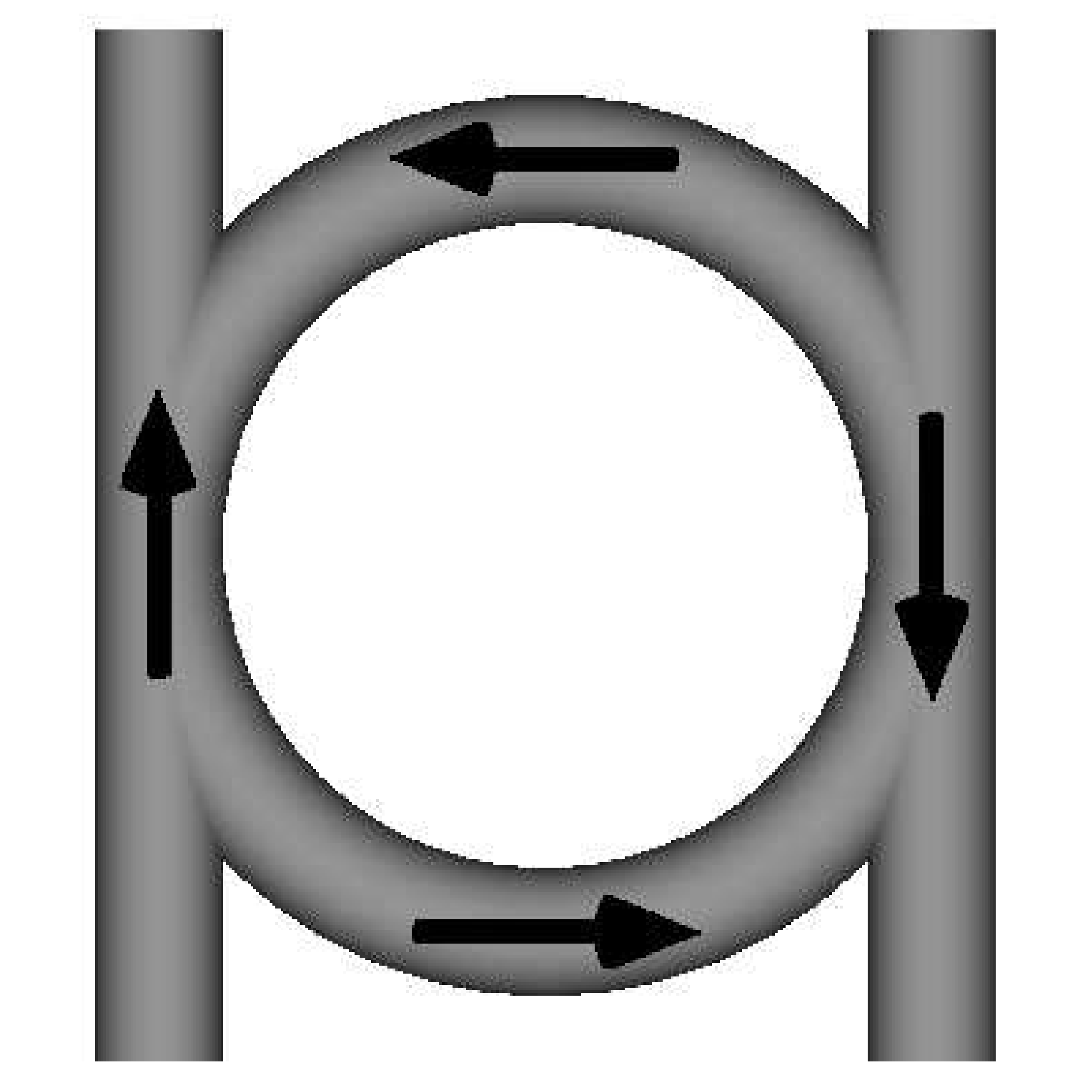}   
      \includegraphics[width=0.24\textwidth]{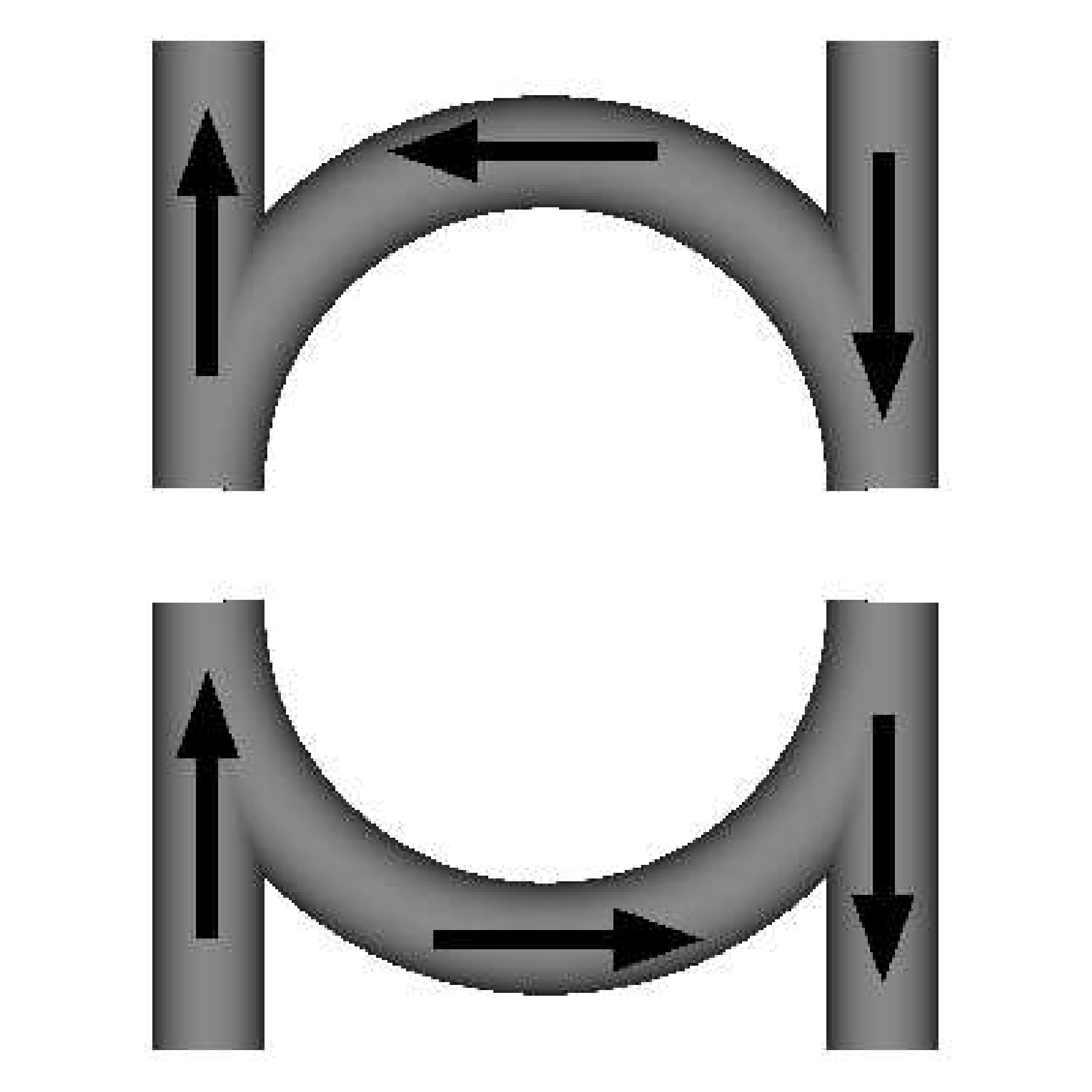}  
     \includegraphics[width=0.24\textwidth]{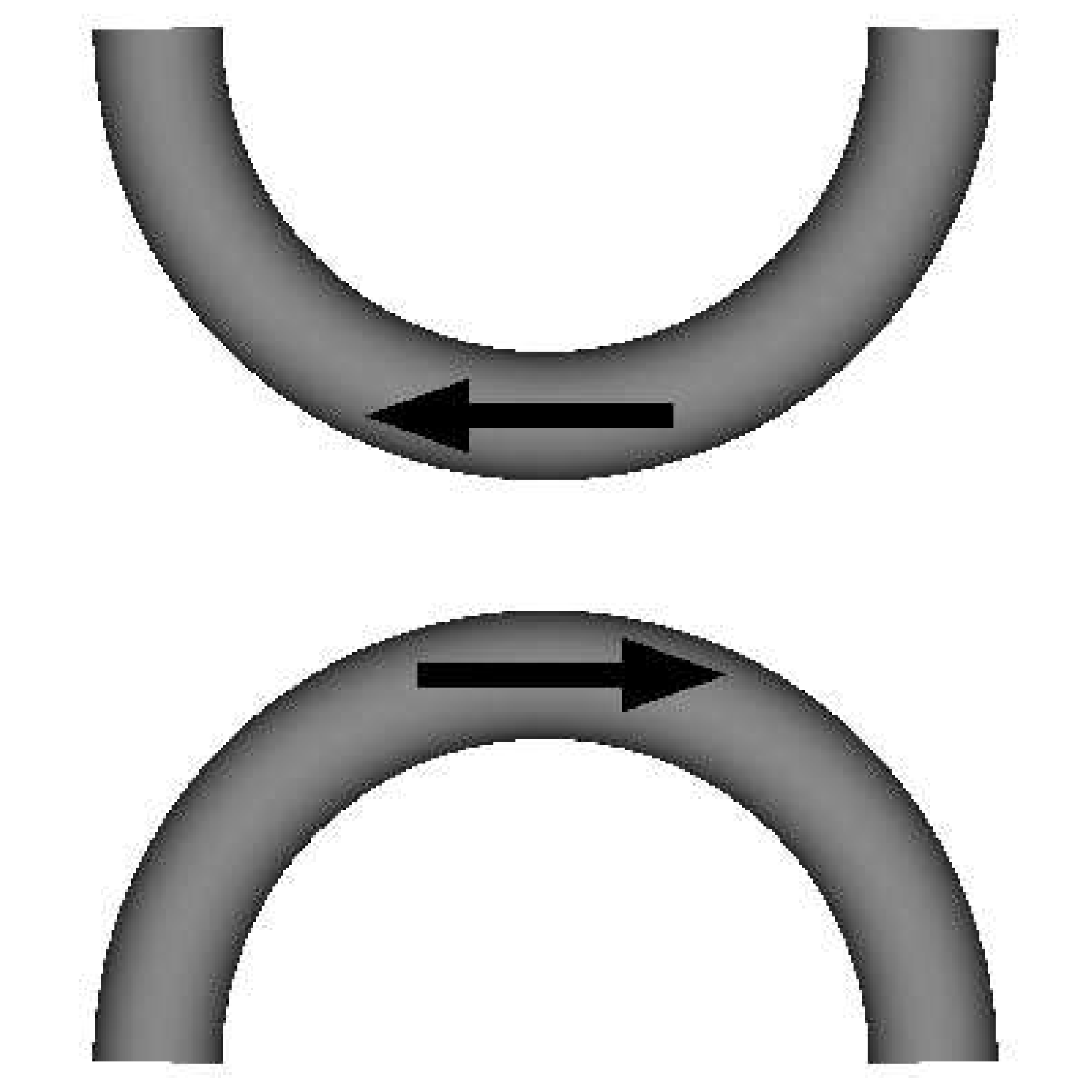} 
   \caption{Cartoon  sequence  illustrating
   the addition of a flux ring to a configuration by a magnetic reconnection process.}
   \label{fig:addstwist}
\end{figure}

Here we also take a simplified approach to consider the local reconnection process, aiming to model
the effect of reconnection on the field topology in a three-dimensional magnetic separator configuration.
As discussed in Section~1, such configurations are thought to be likely sites for current sheet formation and
associated reconnection in the solar atmosphere.
To construct our model we take advantage of a generic feature of reconnection, namely, the presence of an electric field 
with a component parallel to the magnetic field in a localised region.  Such an electric field will, from Faraday's law, 
generate a magnetic flux ring around it.  The flux ring will grow in strength until the reconnection ceases.
This situation is illustrated in Figure~\ref{fig:addstwist} and can be further motivated by considering the equation for the
evolution of magnetic helicity.   Expressing the electric field, $\mathbf{E}$, as
\[ \mathbf{E} = - \frac{\partial \mathbf{A}}{\partial t} - \nabla \phi\] (where $\mathbf{A}$ is the
vector potential for {the magnetic field} $\mathbf{B}$, 
{$\phi$ is a potential and $t$ represents time}),
 we have that the helicity density evolves as
\begin{equation}
\label{eq:helicityevolution}
 \frac{\partial ({\bf A}\cdot {\bf B})}{\partial t} + \nabla \cdot \left(\phi {\bf B} + {\bf E}\times{\bf A}\right)
= - 2 {\bf E}\cdot {\bf B}. 
\end{equation}
In general three-dimensional situations the condition for magnetic reconnection
to occur is the existence of isolated regions at which ${\bf E}\cdot {\bf B} \neq 0$.
Such regions act as source terms for magnetic helicity (see the right-hand side of Equation~\ref{eq:helicityevolution})
thus imparting a localised twist to the configuration at a reconnection site.

We may investigate the effect that reconnection has on a certain magnetic field topology 
by determining the effect of additional localised twist within the configuration.  
Adding any new field component in a manner consistent with Maxwell's equations provides a basis for realistic field modifications.
Although the question of which states could be accessed in a dynamic evolution is outwith the scope of the model,
comparison of results with large-scale numerical simulations may help to determine the plausibility of results
(see Section~\ref{sec:discussion} for a discussion of this point).

Moving on to the specific details of our model, we take as a basic (pre-reconnection) state a
potential magnetic field configuration in which two magnetic null points at $(0,0,\pm z_{0})$ 
are connected by a separator and is
given by 
 \begin{equation}
\label{eq:basicfield}
 {\bf B}_{0} =  \frac{b_{0}}{L^{2}} \left( 
 x\left(z-3z_{0}\right) \ {\hat{{\bf x}}} 
 + y\left(z+3z_{0}\right) \ {\hat{{\bf y}}}
+ \left( z_{0}^{2}-z^{2}+ x^{2}+y^{2} \right) {\hat{{\bf z}}}
\right).
\end{equation}
{Here $z_{0}$ determines the location of the two nulls along the $z$-axis while
$b_{0}$ and $L$ give the characteristic field strength and length scale, respectively.  Here we set}
$z_{0}=5$, $b_{0}=1$, and $L=1$.
  This configuration is exactly that illustrated in Figure~\ref{fig:initialstate2}.
 The spine of the upper null point (itself located at $x=y=0$ $z=5$) is the (one-dimensional) unstable 
 manifold of the null and lies along the line $(0,y,5)$.  The fan surface of the upper null, the (two-dimensional)
 stable manifold lies in the plane $y=0$ and is bounded below by the spine of the lower null-point.
 The spine of this lower null point (itself located at  $x=y=0$, $z=-5$) is the (one-dimensional) stable
 manifold of the null and lies along the line $(x,0,-5)$ while the fan surface is the (two-dimensional) unstable manifold.
 It lies in the plane $x=0$ and is bounded above by the spine of the upper null point.
 The two null points are connected by a separator at the intersection of the two fan surfaces, $x=0, y=0$,
 $z \in [-5, 5]$.  Throughout we consider the magnetic field over the spatial domain $x, y \in [-1,1]$ and $z \in [-5.5,5.5]$.

To simulate the topological effect of reconnection in this configuration we add a magnetic flux ring of the form
\begin{equation}
\label{eq:fluxring}
{\bf B}_{\text{ring}} =  \nabla \times \left( b_{1}  a \exp \left( - \frac{(x-x_{c})^{2}}{a^{2}} - 
\frac{(y-y_{c}^{2})^{2}}{a^{2}} -\frac{(z-z_{c})^{2}}{l^{2}} \right)
{\hat{{\bf z}}} \right).
\end{equation}
Here the flux ring is centred at $(x_{c}, y_{c}, z_{c})$,  the parameter 
$a$ relates to the radius of the ring, $l$ to the height and $b_{1}$ to the field strength.  In the primary model 
considered here we choose to centre the flux ring along the separator (at $x_{c}=y_{c}=z_{c}=0$) and
take the parameters $a=1/2$, $l=1$ and $b_{1}=20$ giving the particular flux ring
\begin{equation}
\label{eq:fluxringmain}
{\bf B}_{1} =  \nabla \times \left( 10 \exp \left( - 4 x^{2} - 4y^{2} - z^{2} \right) {\hat{{\bf z}}} \right).
\end{equation}

We add this flux ring \eqref{eq:fluxringmain} to the potential field \eqref{eq:basicfield}
in a smooth manner, taking a time evolution satisfying Faraday's law,
\[ \nabla \times \mathbf{E} = - \frac{\partial \mathbf{B}}{\partial t}.\]
Specifically we set
\begin{equation}
\label{eq:timeevolution}
{\bf B} = {\bf B}_{0}+ 
{\frac{t}{\tau} \ {\bf B}_{1}}, \ (0 \leq t \leq \tau), \ \ \ \ \  \ \ \ \ \
{\bf E} = - 10 \exp \left( - 4 x^{2} - 4y^{2} - z^{2} \right) {\hat{{\bf z}}}
\end{equation}
with $\tau=1$ so that  the time evolution takes place in $0 \leq t \leq 1$.
(Note that the gradient of a scalar could also be added to the electric field which could allow for the superimposition 
of a stationary ideal flow.  We have, for simplicity, neglected this possibility.)
Similar evolutions can be obtained for the addition of the more general flux ring \eqref{eq:fluxring}
to the potential field \eqref{eq:basicfield}.

The addition of the magnetic flux ring \eqref{eq:fluxringmain}  creates a localised region of twist in the centre of the domain
and we now wish to determine whether and how the magnetic field topology is changed.
In order to do so first notice that the flux ring is sufficiently localised that the magnetic field at the null points
remains unchanged during the time evolution.  Accordingly for each magnetic null point we may trace magnetic 
field lines in the fan surface in the neighbourhood of the null point and out into the volume.
This method allows us to determine how the fan surfaces are deformed by the reconnection  and to 
locate any intersections of these surfaces, i.e.~separators.  We describe our findings on the field topology in the following section.

\section{A bifurcation of separators}
\label{sec:bif}

\begin{figure}[htbp] 
 \centering
 \includegraphics[width=0.26\textwidth]{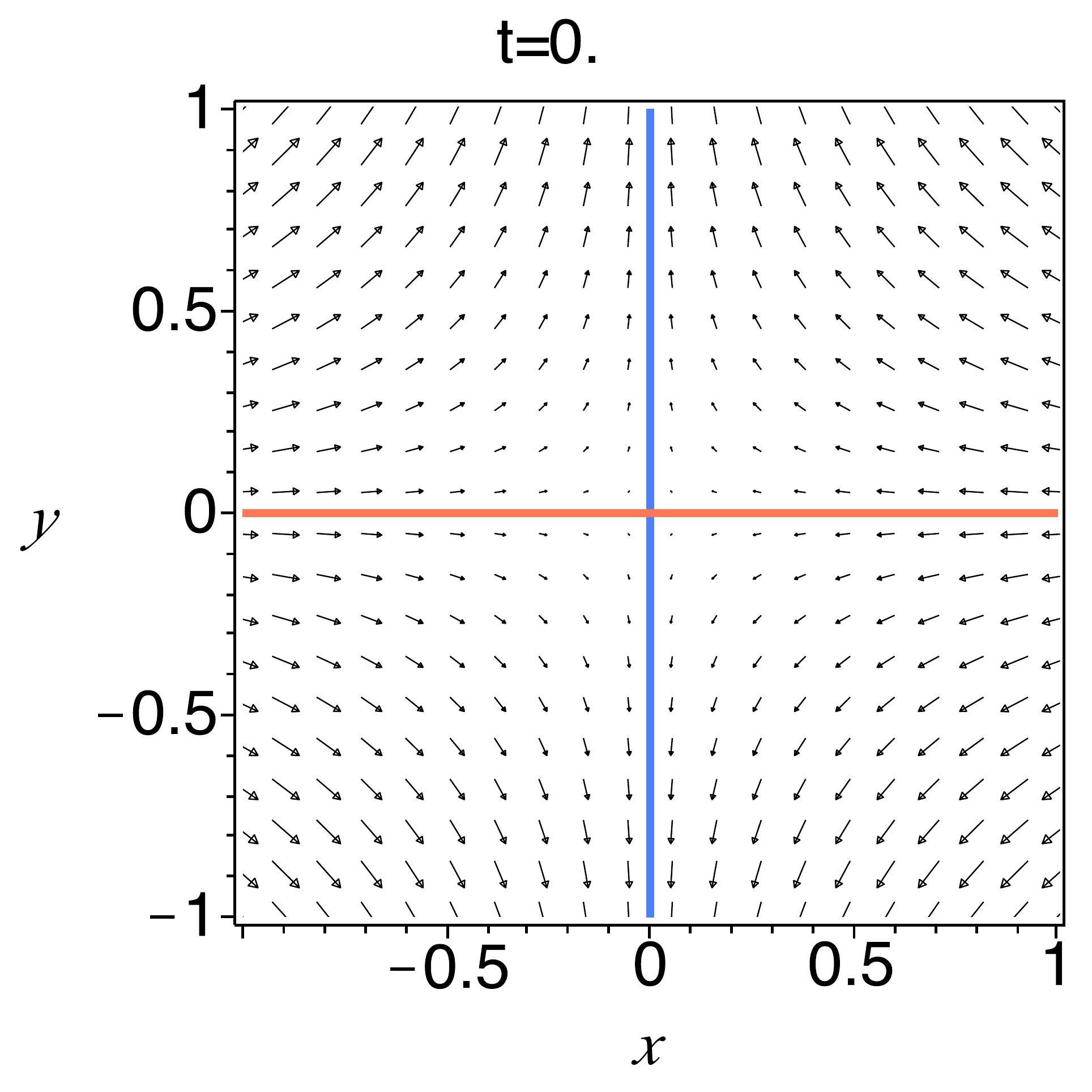} 
 \includegraphics[width=0.26\textwidth]{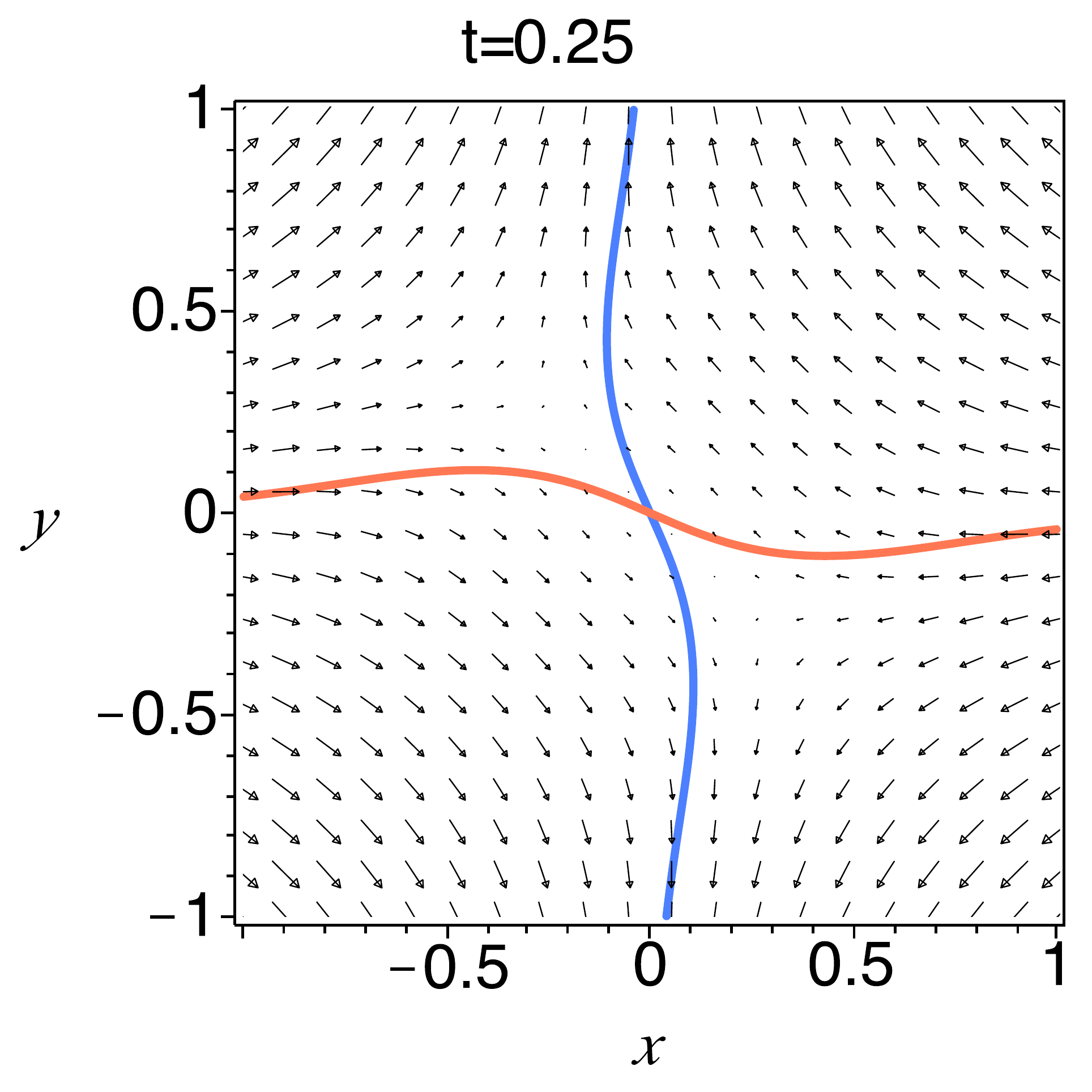} 
  \includegraphics[width=0.26\textwidth]{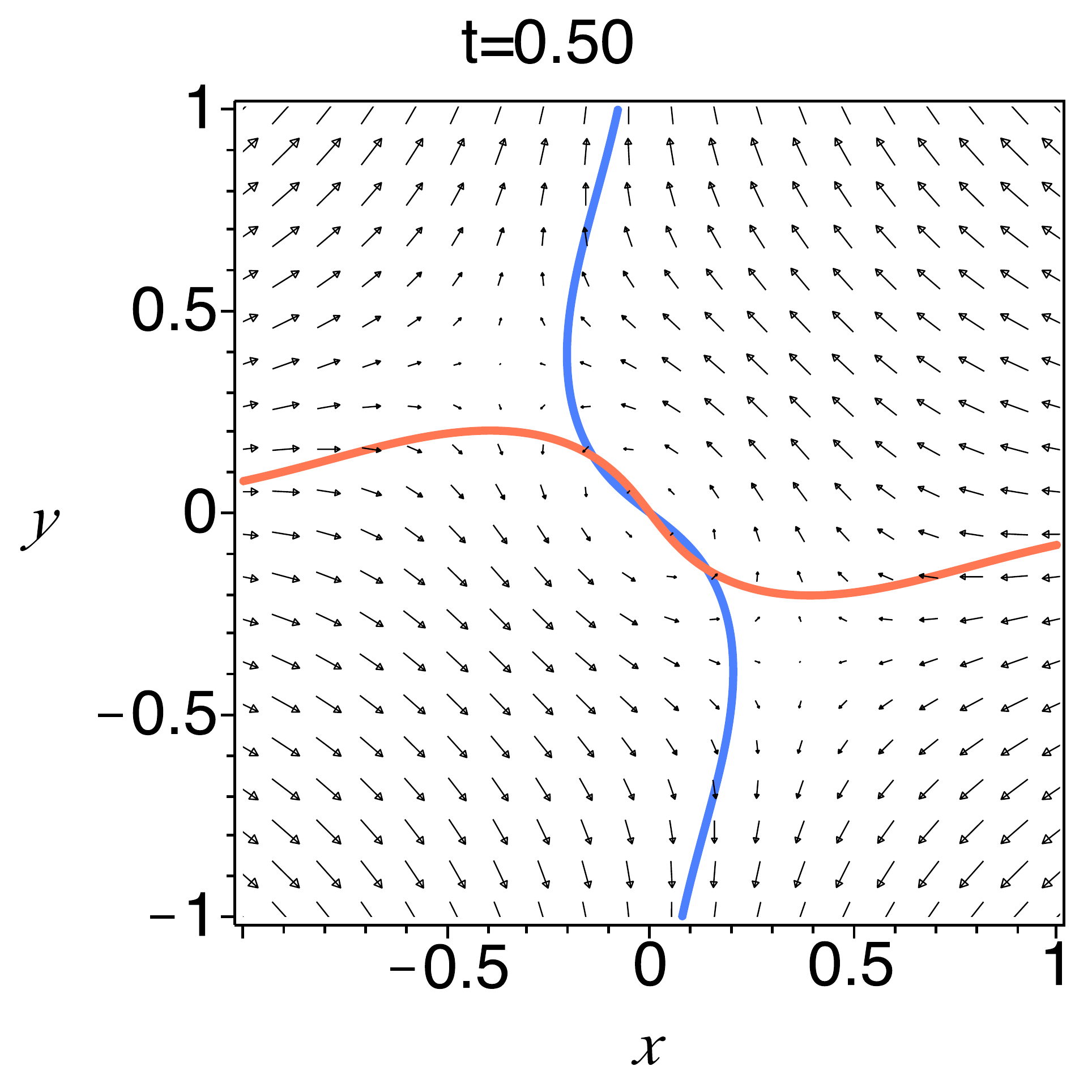}  
  \includegraphics[width=0.26\textwidth]{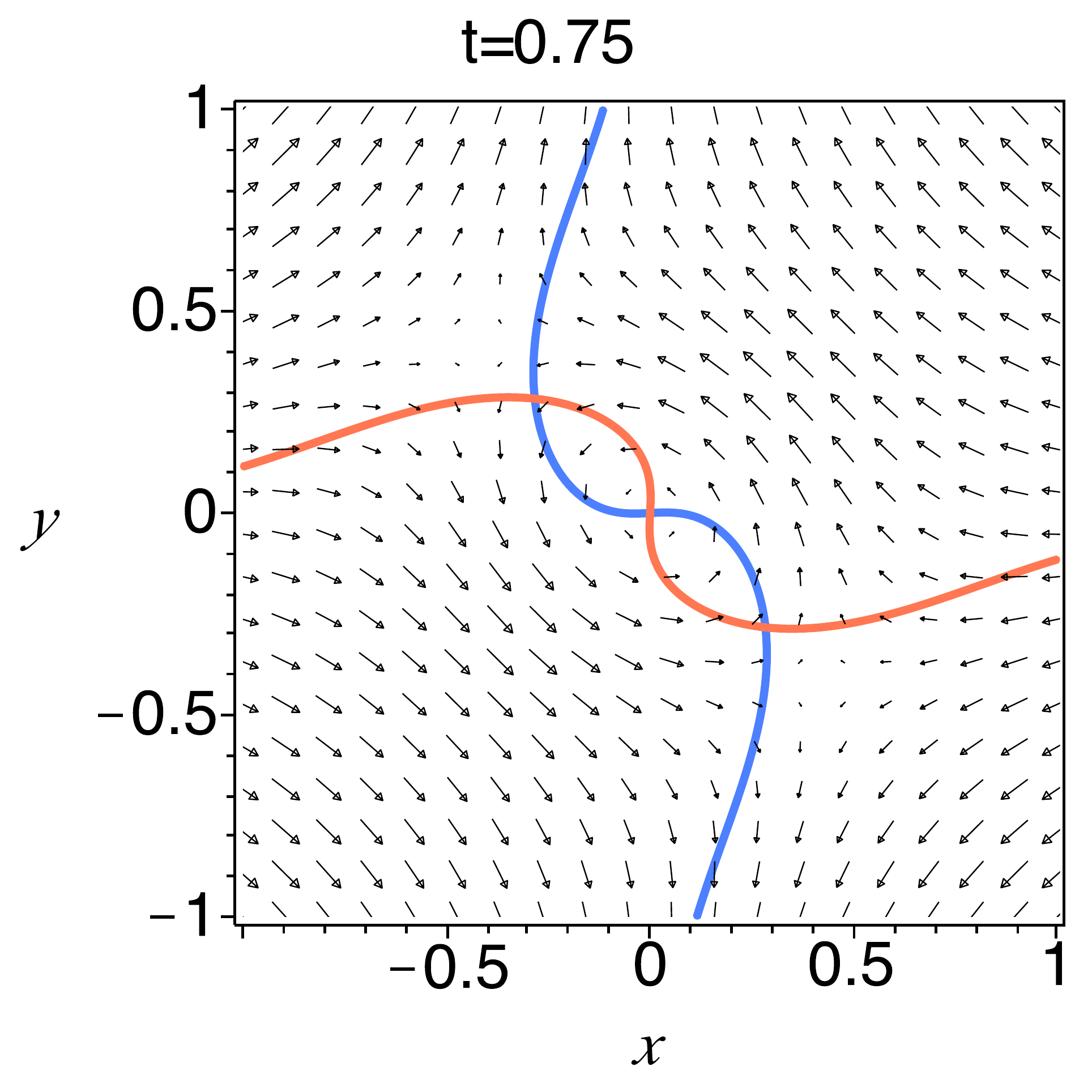} 
\includegraphics[width=0.248\textwidth]{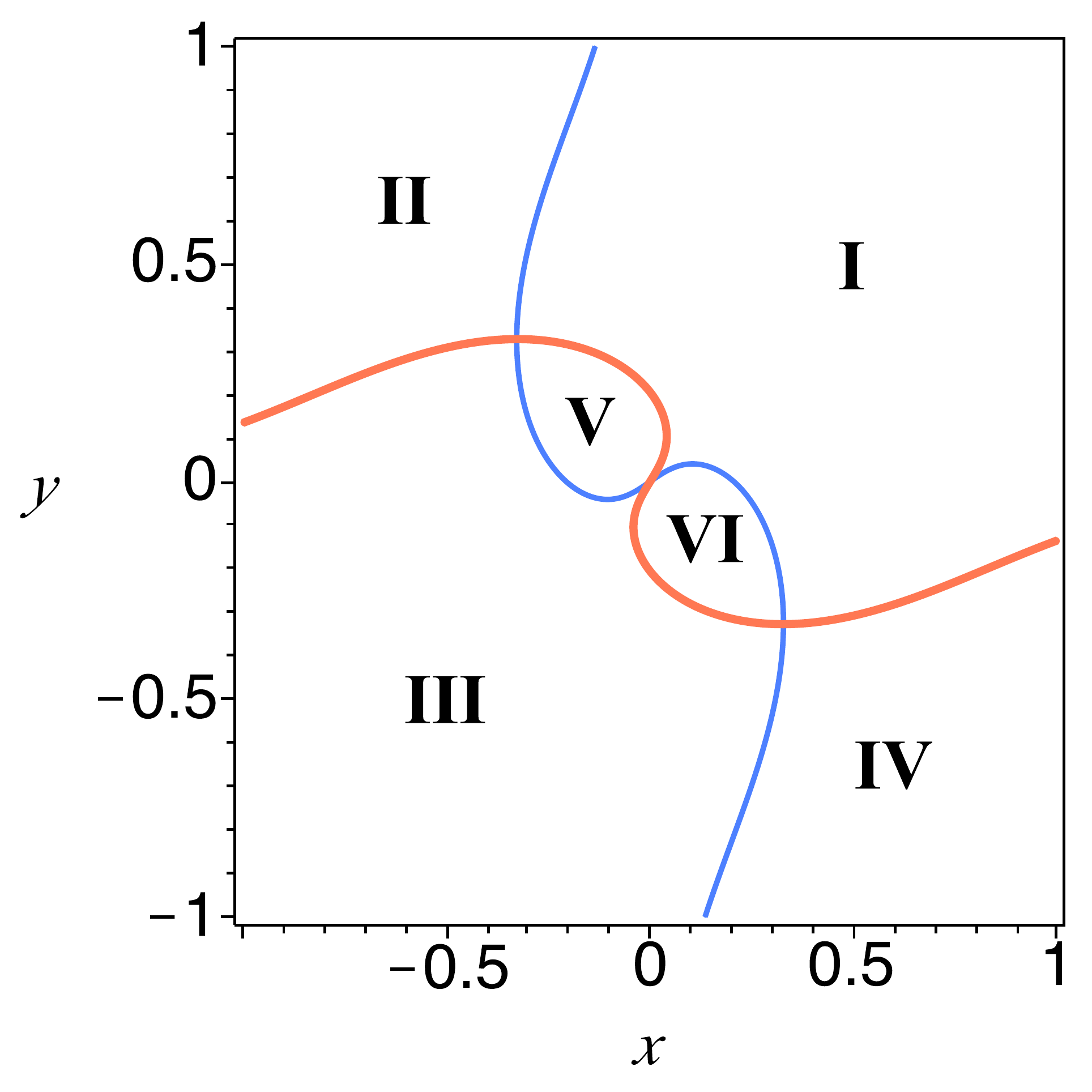} 
   \caption{Intersection of the fan surfaces of the lower (blue) and upper (orange) nulls with the $z=0$ plane
  at times indicated.  Superimposed is the vector field $(B_{x},B_{y})$ of horizontal magnetic 
  field components at the relevant times.  A folding of the fan surfaces creates new separators in pairs
  as the fan surfaces intersect.  The lower-right--hand image indicates our labelling of the topologically
  distinct regions.}
   \label{fig:fansz0}
\end{figure}

In order to follow the evolution of magnetic field topology during the reconnection we begin by showing 
in Figure~\ref{fig:fansz0} the intersection of the fan surfaces with the $z=0$ plane.  The strength of the flux ring 
increases linearly in time up to the final state which we have normalised as $t=1$.
We show the intersections  by tracing field lines from each fan surface in the close neighbourhood of the 
corresponding null which is possible
since the disturbance is localised near the centre of the domain and so the eigenvalues associated with the nulls
do not change in time.  In the images the fan surface of the lower null is coloured blue and that of the upper null
in orange.

We first note that in the initial phase of the process the angle between the fan surfaces decreases.  If the
reconnection is weak the process can stop in this phase without leading to any change in the
magnetic skeleton.  However, a stronger reconnection event can lead to a further closing of the angle between
the fan surfaces until they intersect (at about $t=0.44$ in our model).
Recall that crossings of the fan surfaces give the location of magnetic separators in that particular plane. 
Hence the process creates two new separators and correspondingly two new magnetic flux domains. 
In order to properly identify the various flux domains we label each of the distinct topological regions 
with the numbers I--VI, as shown in the lower-right--hand image of Figure~\ref{fig:fansz0}.

We shall examine the nature of these flux domains
 later in this same section, but at this point it is already possible to make some general statements about this type of bifurcation. 
Obviously the manner in which the fan planes can fold and intersect leads to the
 process always creating (or the reverse process annihilating) separators in pairs. There is no way in which we can create a single new separator as long as the reconnection  is localised. (This excludes that the whole domain under consideration is non-ideal and that separators enter or leave the domain across the boundary).

Also shown in Figure~\ref{fig:fansz0} are components of the vector field $(B_{x},B_{y})$ in the $z=0$ plane
{which is the vector field perpendicular to the central separator at $(0,0,0)$.  
This field structure is initially hyperbolic but becomes elliptic as the reconnection   continues.
 Additionally a separator typically has both elliptic and hyperbolic field regions along its length.}
These findings coincide with those of Parnell {\it et al.} (2010) who examined the local magnetic
field structure along separators in a three-dimensional MHD simulation and found both elliptic and hyperbolic
perpendicular field components.

\begin{figure}[htbp] 
\centering
\includegraphics[width=0.35\textwidth]{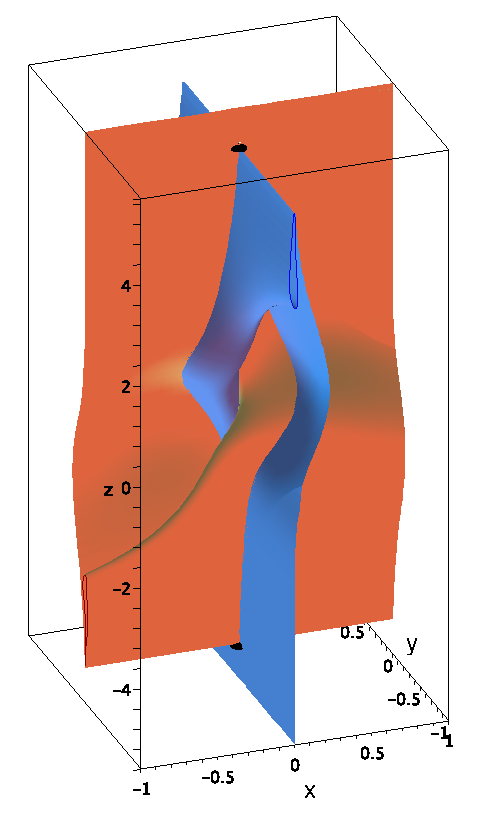} 
 \includegraphics[height=10cm,width=0.32\textwidth]{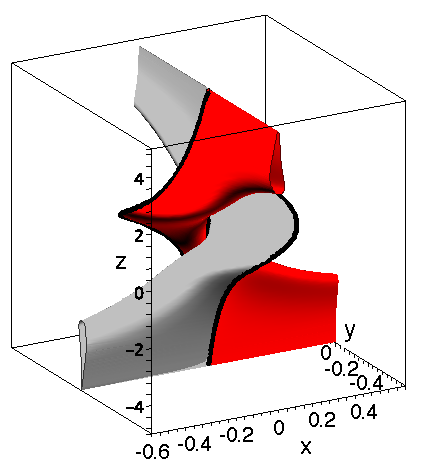} 
   \caption{({\it Left}) Fan surfaces of the lower (blue) and upper (orange) nulls at $t=0.7$.  
Three magnetic separators are present where the fan surfaces intersect.
The location of the nulls is marked with black spheres.
The folding and intersection of the fan surfaces has created two new flux domains.
The corresponding flux tubes are also shown separately ({\it right})  where the three separators are 
marked with black lines (the aspect ratio of the image has been further distorted to allow the main features
to be identified).}
   \label{fig:fans07}
\end{figure}

A three-dimensional view of the fan surfaces at $t=0.7$ (an instant when three separators are present) is shown in 
Figure~\ref{fig:fans07}.  
Since each separator begins and ends at the null points, the surfaces must fold in a complex way 
such that the three intersections merge and coincide at the nulls.
This folding takes place as the fan surface of one null approaches the spine of the other null and
creates `pockets' of magnetic flux running parallel to the spine, as shown in Figure~\ref{fig:fans07} where
the flux passing through the $y=-1$ boundary can be seen.
The new flux domains alone are shown separately in the right-hand image of Figure~\ref{fig:fans07} 
and correspond to the two new flux domains (types V and VI) that appear in
the first bifurcation as shown in Figure~\ref{fig:fansz0}.

We now examine in more detail the connectivity of the magnetic flux in the topologically 
distinct flux domains I--VI (Figure~\ref{fig:fansz0}), focusing on the flux that passes through the central plane, 
$z=0$ (flux above the upper null and below the lower null is not of interest here).
The flux through the central plane can be distinguished in two ways. 
Firstly we can follow field lines in the negative direction to determine on which side of the 
fan plane of the lower null they end ($x>0$ or $x<0$) and secondly
 we can follow the field lines in positive direction to determine on which side of the fan plane of the upper null they 
 end ($y>0$ or $y<0$).   Combined this technique shows the nature of the field line connectivity in each 
 of the four different flux domains.
  Flux in the domain labelled I enters the domain through the boundary $x=+1$ and leaves through the boundary
$y=+1$.  Flux in domain II enters through $x=-1$ and leaves through $y=+1$, that in domain III enters through
$x=-1$ and leaves through $y=-1$ while that in domain IV enters through $x=+1$ and leaves through $y=-1$.  
Note that this distinction of fluxes is based on topological features of the null points and is independent of the 
particular choice of our boundary.

Following the first bifurcation at $t\approx 0.440$ the two new flux domains V and VI are created. 
Field lines in flux domain V have the same
 basic connectivity type as those in flux domain IV (i.e.~they enter through
 through the $x=+1$ boundary and leave through $y=-1$ boundary).   
  Field lines in flux domain VI have
  the same basic connectivity type as those in flux domain II (i.e.~they enter
 through the $x=-1$ boundary and leave through $y=+1$ boundary).
 Although these two basic connectivity types existed prior to the bifurcation, the flux is enclosed within 
 topologically distinct regions as it threads the new `pockets' previously described (see Figure~\ref{fig:fans07}). 
  
\begin{figure}[htbp]
 \centering
  \includegraphics[width=0.425\textwidth]{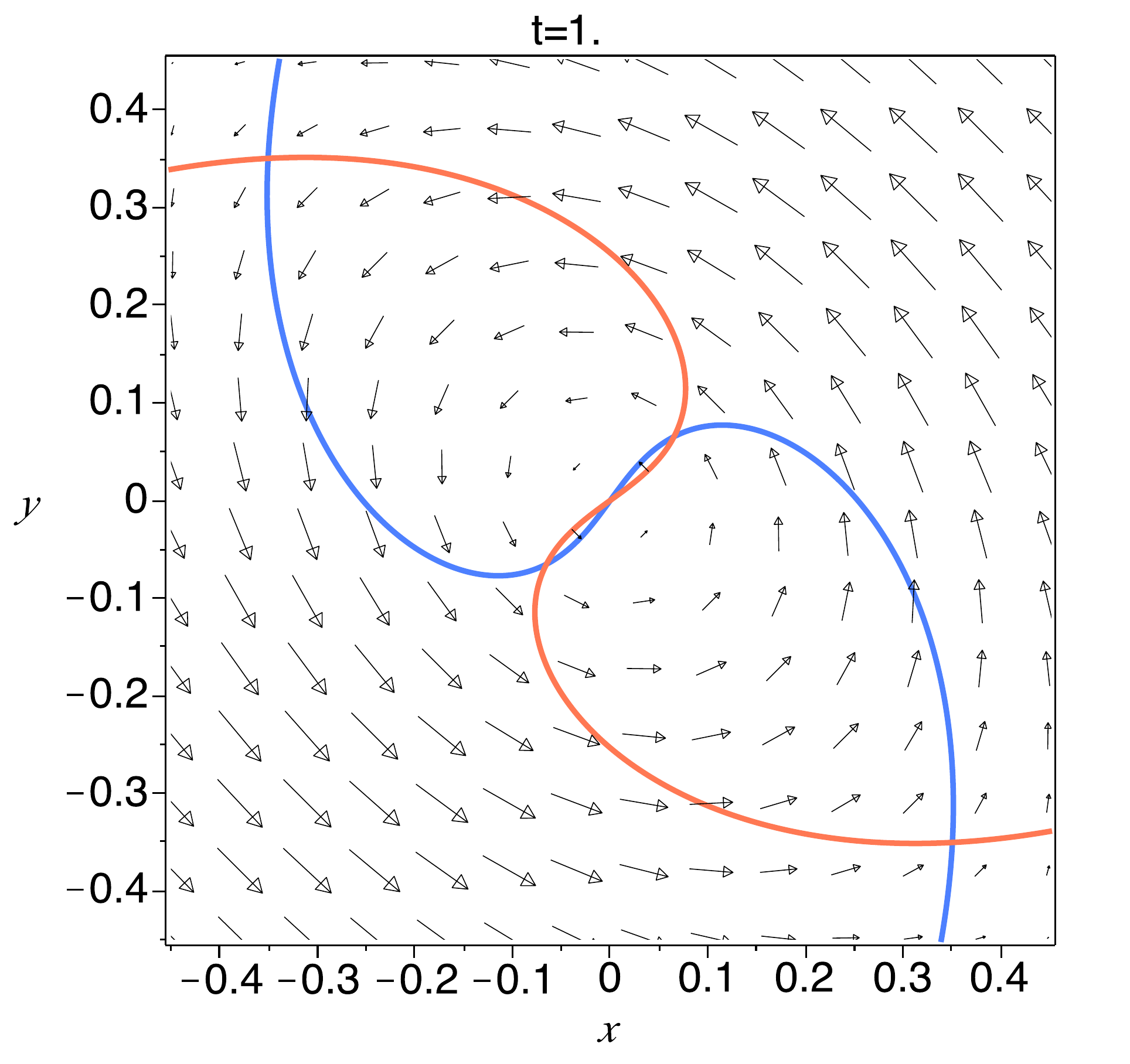} 
     \includegraphics[width=0.4\textwidth]{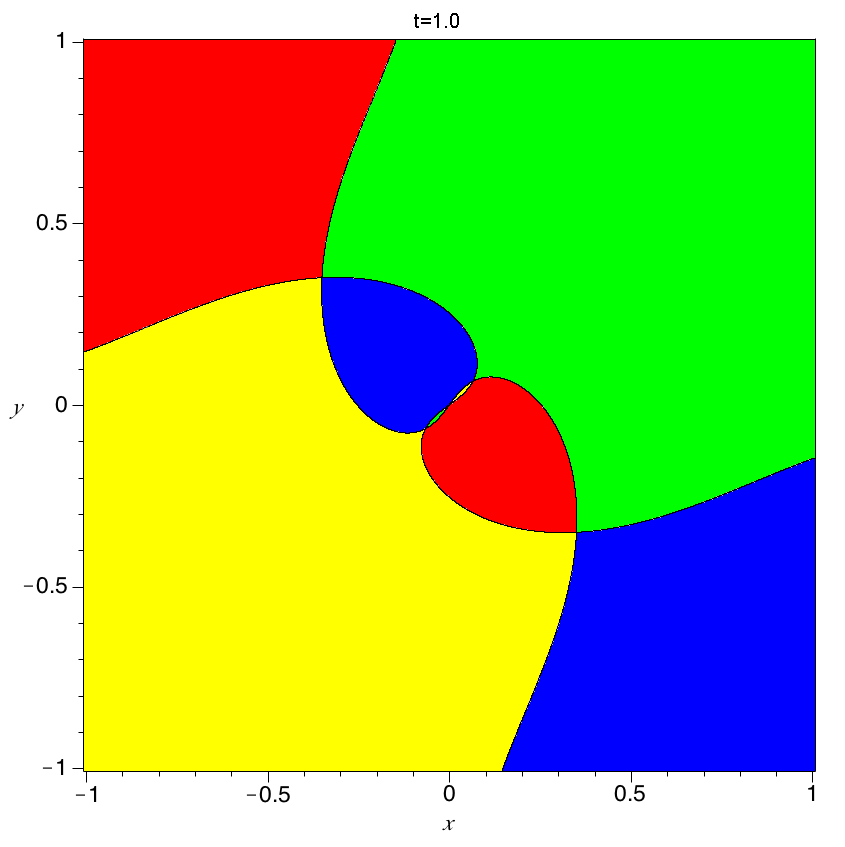} 
   \caption{  ({\it Left}) Intersection of the fan surfaces of the lower (blue) and upper (orange) nulls with the $z=0$ plane
   at $t=1$.  Superimposed is the vector field $(B_{x},B_{y})$.  Five separators are present at this time.   
 Were the reconnection process to continue, further folding of the fan surfaces would
continue to generate separators in pairs in this manner.
({\it Right}) 
At $t=1$ the various topologically flux domains are coloured according to their boundary connectivity types (for 
the main regions compare with Figure~3) showing the relation between connectivities of the various domains.}
   \label{fig:fansend}
\end{figure}

 As the reconnection   continues, further distortion of the surfaces gives another bifurcation at  
$t \approx 0.957$ generating an additional pair of separators, as shown in Figure~\ref{fig:fansend}.  Thus at
the end of the reconnection   considered here
five separators are present in the configuration, each connecting the same pair of nulls. 
Figure~\ref{fig:fansend} also shows
a colour coding of the various flux domains according to connectivity type.  This is shown at the end of the
reconnection   and gives a summary of the way the flux domains relate to each other (i.e.~field lines in 
flux domains shown in the same colour  have the same connectivity type).

We envisage this  type of reconnection  as occurring, for example, at the dayside
magnetopause due to interaction with the interplanetary magnetic field.
Accordingly, a crucial question is exactly how the changes in flux connectivity occur and how 
flux is transferred between topologically distinct domains.
In the magnetopause example this would have implications for
how the solar wind plasma can interact with that in the magnetosphere.
We begin to address this question in the following section.

\section{Evolution of Magnetic Flux}
\label{sec:evolve}

In order to track the magnetic flux in time and determine how the changes in connectivity occur we examine
some particular {\it magnetic flux velocities} and so begin by briefly discussing their motivation.
Recall that an ideal evolution of the magnetic field is one satisfying
\[ \mathbf{E} + \mathbf{v} \times \mathbf{B} = \mathbf{0}, \]
{(where ${\bf v}$ is the plasma velocity)}
the curl of which gives
\begin{equation}
\label{eq:induct}
 \frac{\partial \mathbf{B}}{\partial t}  - \nabla \times (\mathbf{v} \times \mathbf{B}) = \mathbf{0}.
 \end{equation}
In such a situation the magnetic field and a line element have same evolution equation
and so flux is `frozen-in' to the plasma and the magnetic topology is conserved.

A real plasma evolution has 
\begin{equation}
\label{eq:real}
 \mathbf{E} + \mathbf{v} \times \mathbf{B} = \mathbf{N} 
 \end{equation}
where $\mathbf{N}$ represents some non-ideal term 
(such as $\eta \mathbf{J}$ in a resistive MHD evolution
{where $\eta$  is the resistivity and ${\bf J}$ the electric current}) 
which is typically localised to some region of space.
In this paper the effect of the non-ideal term is modelled by the addition of a magnetic flux ring.
Even with the inclusion of the non-ideal term 
we may sometimes still find a velocity $\mathbf{w}$ with respect to which the magnetic flux is frozen-in
if \eqref{eq:real} can be written as
\begin{equation}
\label{eq:modreal}
 \mathbf{E} + \mathbf{w} \times \mathbf{B} = \nabla \Phi,
 \end{equation}
 where $\Phi$ is an arbitrary function, since taking the curl again and using Faraday's law gives an equation of the form
 \eqref{eq:induct}.

In three-dimensional magnetic reconnection we have localised regions where $\mathbf{E} \cdot \mathbf{B} \neq \mathbf{0}$
and, as a result, a unique flux velocity $\mathbf{w}$  cannot be found (Hornig \& Priest 2003).
Instead we may consider field lines as they enter and leave the non-ideal ($\mathbf{E} \cdot \mathbf{B} \neq \mathbf{0}$)
region, fixing them on one of these sides.  
{We choose a transversal surface that lies below the non-ideal region and integrate along 
magnetic field lines into and out of the non-ideal region. 
Parameterising a magnetic field line by ${\bf x}(s)$  and starting the integration from the point 
${\bf x}_{0} = {\bf x}(0)$  (with $\Phi({\bf x_{0}}) = 0$) on 
the transversal surface we may determine $\Phi$ as}
\[ \Phi(\mathbf{x(s)}) := \int_{s'=0}^{s} \mathbf{E}_{\parallel} ({\bf x}(s')) \ ds',
\ \ \ \ \frac{\textrm{d} {\bf x}(s)}{
\textrm{d}s} = \frac{{\bf B}}{\vert {\bf B} \vert},
 \]
and subsequently $\mathbf{w}$ as 
\[ \mathbf{w} = \frac{\mathbf{B} \times \left(\nabla \Phi - \mathbf{E}\right)}{B^{2}}. \]

Carrying out the integration for $\Phi$ until field lines meet a boundary and leave the domain and noting that
$\mathbf{E} = \mathbf{0}$ at these boundaries we can find a particular flux velocity
\begin{equation}
\label{eq:ourw}
\mathbf{w}^{*} = \frac{\mathbf{B} \times \nabla \Phi }{B^{2}}
\end{equation}
and visualise the effect of reconnection on the magnetic flux by showing
its component perpendicular to the boundaries.
This procedure can only be carried out in the presence of a single magnetic separator; multiple separators lead to
ambiguities in the potential $\Phi$.
Note that in the example considered here the 
electric field which is responsible for the
non-ideal evolution decays exponentially away from its centre.  At $z=-2.5$
the electric field has fallen to a value of $0.2\%$ of the maximum in the domain,
sufficiently small to be considered an ideal environment and so we choose this as 
 our transversal surface for the calculation of $\Phi$.

\begin{figure}[htbp] 
   \centering
      \includegraphics[width=0.2\textwidth]{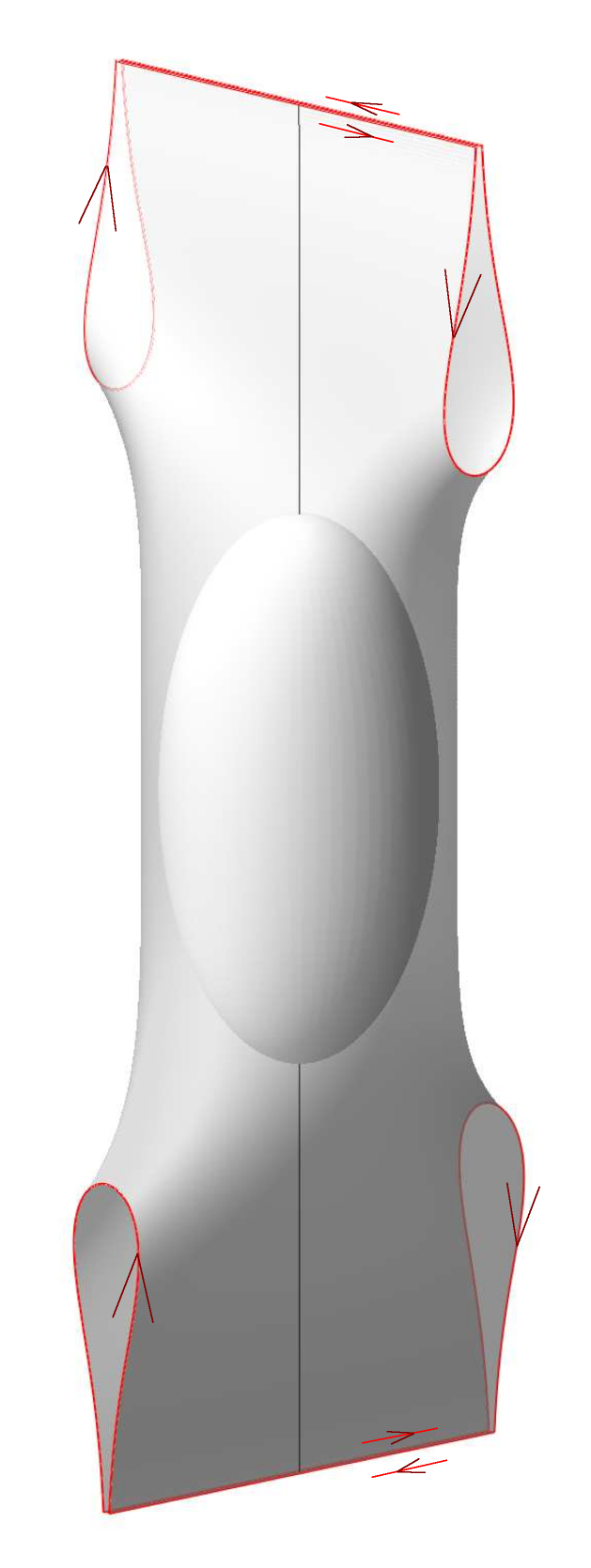} 
      \ \ \ \ \ \ \       \ \ \ \ \ \ \ 
            \includegraphics[width=0.19\textwidth]{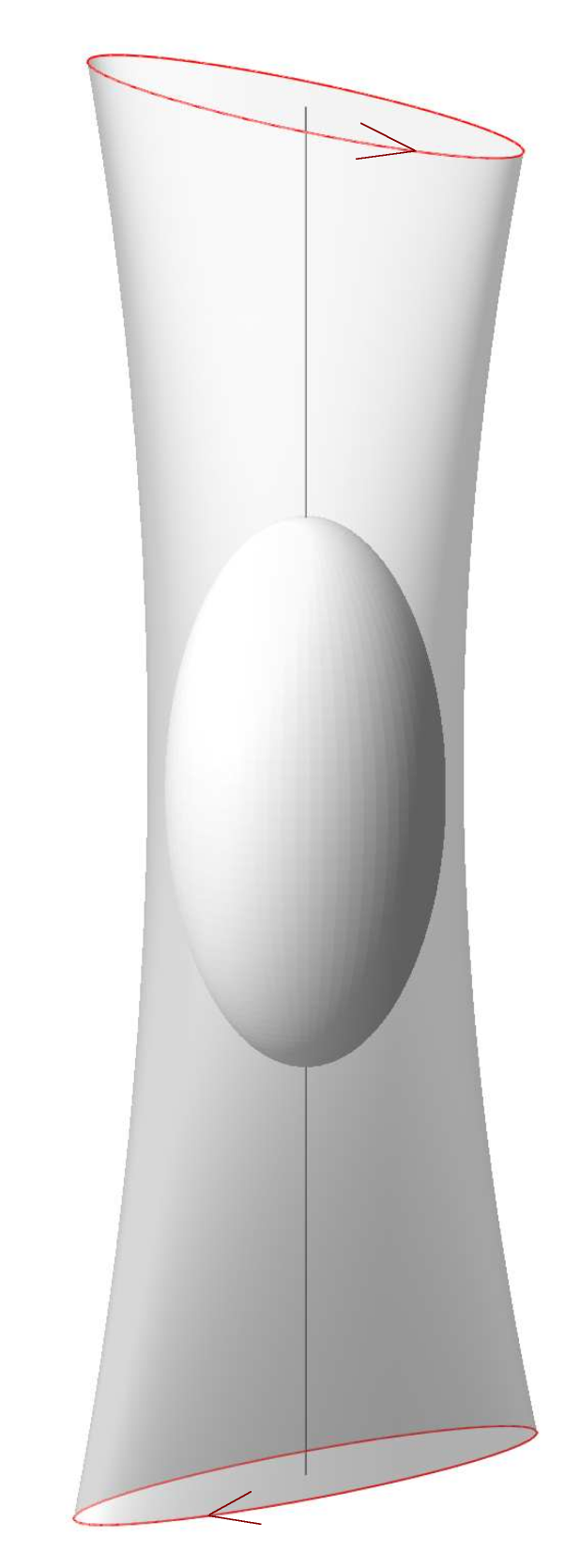} 
   \caption{Cartoon illustrating the nature of the magnetic flux velocities in reconnection.  The left-hand
   image shows the separator case while the right-hand image shows the non-null case.}
   \label{fig:sephft2}
\end{figure}

\begin{figure}[htbp] 
   \centering
      \includegraphics[width=0.7\textwidth]{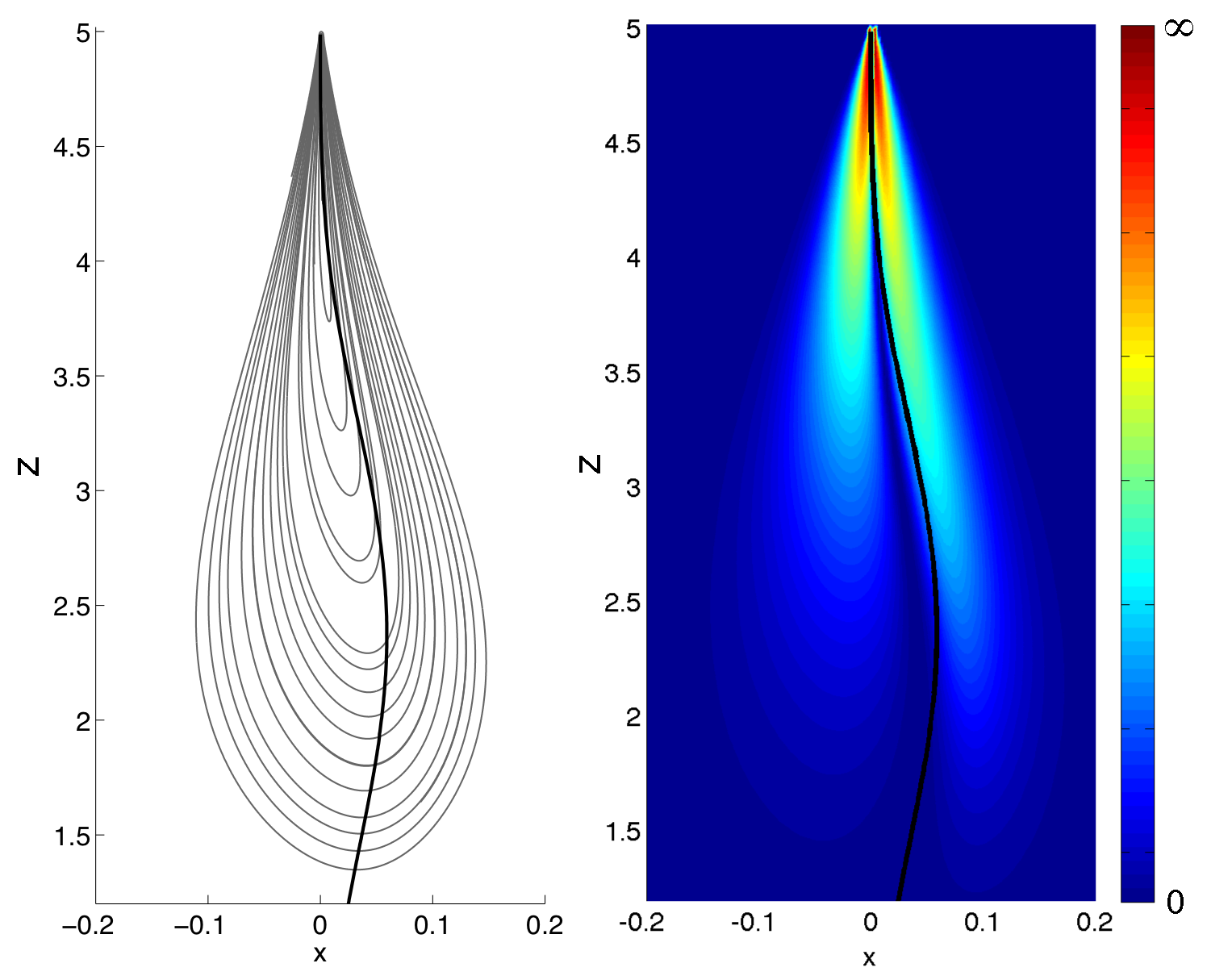} 
   \caption{The magnetic flux velocity on the $y=-1$ boundary at $t=0.25$.  The left-hand image
   shows streamlines of the flux velocity (grey lines) and the location of the fan surface of the lower null (black line).
   The right-hand image indicates the magnitude of the flow which becomes singular at the spine ($x=0, z=5$).
   The direction of the flow is anti-clockwise.   
   }
   \label{fig:fluxvel}
\end{figure}

In the cartoon in Figure~\ref{fig:sephft2} (left-hand image) the direction of the flux velocity 
${\bf w}^*$ is shown for a surface of field lines bounding the non-ideal region. 
The situation considered here  can be compared with that without null points (Hornig \& Priest 2003) 
which is indicated in Figure~\ref{fig:sephft2} (right-hand image).  In the non-null case the flux velocity has the well-known 
feature of counter-rotating flows. With the addition of null points, the cross-section of the flux tube which bounds the 
non-ideal region is split into two separate domains connected by a singular line, the spine, for each of the 
two nulls. Along the spine the flux velocity becomes infinite.  This feature can be also found in 
Figure~\ref{fig:fluxvel}, which shows the component of $\mathbf{w}^{*} $ perpendicular to the $y=-1$ 
boundary at $t=0.25$. The left-hand image shows streamlines of $\mathbf{w}^{*} $ (grey lines) and the 
intersection of the 
fan surface of the lower null with the boundary (black line).  The flow direction is counter-clockwise.
The magnitude of $\mathbf{w}^{*}$ is shown in the right-hand image.  
A logarithmic scale is taken and the flow becomes infinite at the point of intersection of the upper null's spine
($x=0, z=5$).  
The flux velocity shown in  Figure~\ref{fig:fluxvel} is typical of that in the early evolution when only one separator 
is present.  A rotational flux velocity is found within the flux tube passing through the non-ideal region.  Additionally, the 
fan surface behaves as if advected by this flux velocity.

\begin{figure}[htbp] 
\centering
\includegraphics[width=0.19\textwidth]{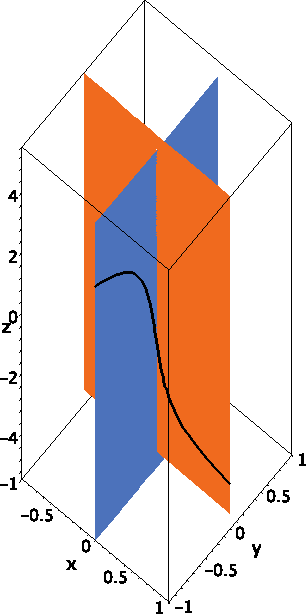} 
\includegraphics[width=0.19\textwidth]{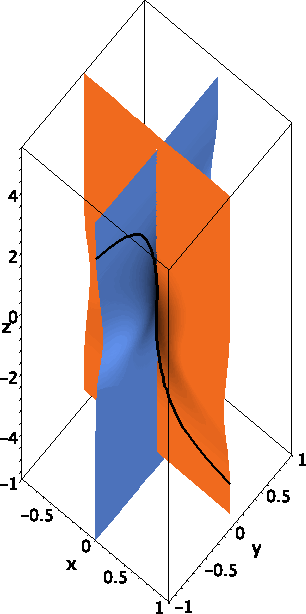} 
\includegraphics[width=0.19\textwidth]{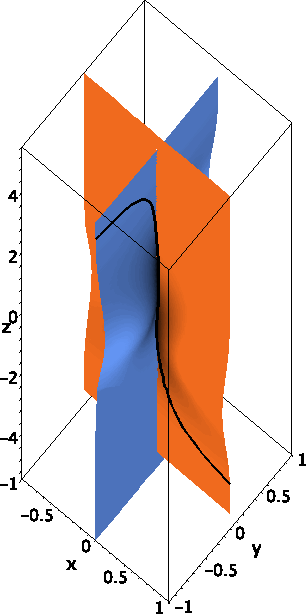} 
\includegraphics[width=0.19\textwidth]{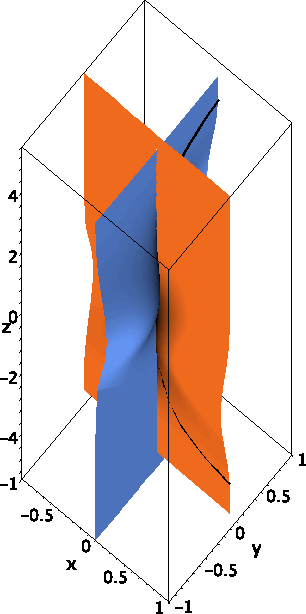} 
\includegraphics[width=0.19\textwidth]{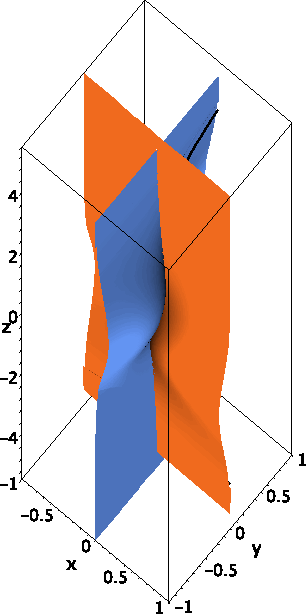} 

\

\

\includegraphics[width=0.19\textwidth]{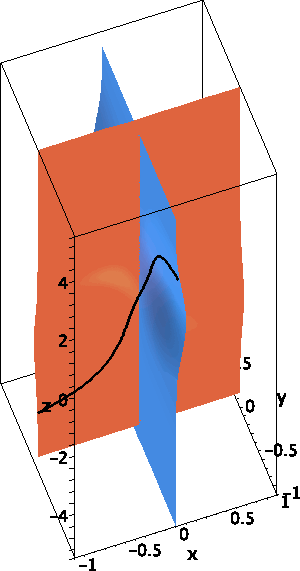} 
\includegraphics[width=0.19\textwidth]{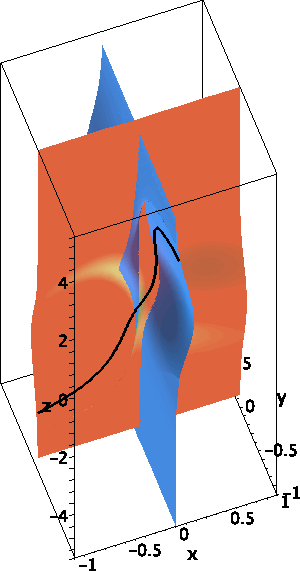} 
\includegraphics[width=0.19\textwidth]{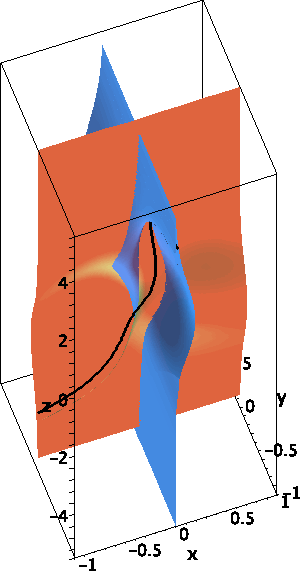} 
\includegraphics[width=0.19\textwidth]{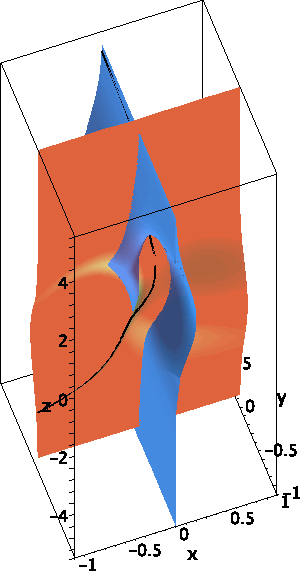} 
\includegraphics[width=0.19\textwidth]{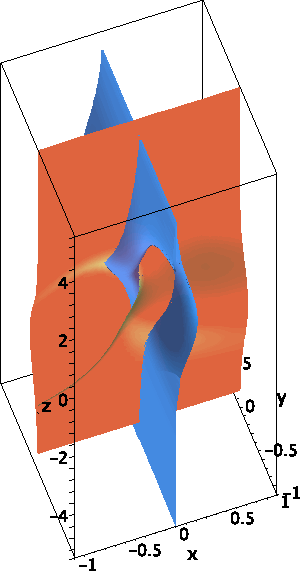} 
\caption{Field line sequences showing the nature of reconnection  in a configuration with 
a single separator (upper panel) and with multiple separators (lower panel).  The
detail of the field line evolution is described in the main text.  
Specific times for the particular sequences shown here are $t=0, 0.25, 0.325, 0.4, 0.55$ (upper panel) 
and $t=0.25, 0.5, 0.55, 0.6, 0.65$ (lower panel).
}
   \label{fig:recntypes}
\end{figure}

With these considerations in mind we may consider the nature of the reconnection   to 
be as follows.  For exactness in the discussion we assume the field lines are fixed below the non-ideal region
although a symmetric situation occurs in the alternative case with field lines fixed above the non-ideal region.
A magnetic field line threading the non-ideal region leaves the domain through a side boundary, $y=-1$, say (that for
which the flux velocity is illustrated in Figure~~\ref{fig:fluxvel}).  
There are two distinct ways the field line topology changes, illustrative phases of which
are shown in Figure~\ref{fig:recntypes}:

\begin{enumerate}
\item
The first situation is shown in the upper panel of Figure~\ref{fig:recntypes}.
The field line initially lies to the right of the fan surface of the lower null (flux domain IV).
The motion of this unanchored end is towards the spine of the upper null.  It reaches the spine in a 
singular moment at which the flux velocity is infinite. Here the field line is connected to the upper null, 
lying in the fan surface of that  null (this is a dynamic situation with the fan surfaces moving in time; the 
fan sweeps across the anchored end of the field line).  The field line then moves through the spine and 
is connected to the opposite side boundary as it leaves the domain (flux domain I).  
In this process reconnection is occurring continuously while the field line is connected to the non-ideal region.
The `flipping' of of the magnetic field line and change in its topology occurs at a singular moment during 
the reconnection. 
\item
The second situation is shown in the lower panel of Figure~\ref{fig:recntypes}.
The field line initially lies to the left of the fan surface of the lower null (flux domain III).
The motion of the unanchored end is again towards the spine of the upper null.  
The  fan surface of the upper null is moving away from the anchored end of the field line
while order to change its global connectivity (pass through the upper null) the entire field line must be lying 
in the fan plane of that null.
Accordingly a `flipping' of the field line can only occur at or after the first bifurcation of separators.
In this bifurcation a fold (pocket) in the fan surface is created which 
sweeps up and over the anchored end of the field line.  
In the moment at which the global field line connectivity changes the field line is connected to the (rising) 
pocket of the fan and the upper null.  In the following instant the unanchored end of the field line 
moves to the opposite side boundary ($y=+1$) and the field connectivity is the new type VI.
\end{enumerate}

While the separator reconnection configuration is sometimes viewed as a three-dimensional analogue
of  two-dimensional reconnection case at an X-type null point, this example illustrates that
the behaviour of the two is, in many ways, quite distinct.
In the two-dimensional case an analysis of the magnetic flux 
velocities (see, for example, Hornig 2007) shows that the reconnection of field lines occurs {\it at}  the magnetic null point 
where the magnetic flux velocity is infinite and the field lines are `cut' and rejoined. 
  Our analysis demonstrates that in  the separator case a magnetic 
null is also key to the reconnection   with field lines passing through the null at the moment their
global connectivity changes, again when the flux velocity is infinite.
However, this is a non-local process with the null itself far removed from the reconnection site (indeed the null
may lie in an ideal environment).  The magnetic separator itself has only an indirect role in the process.
We suggest that, physically, the locations of the singularity in the flux velocity may 
be associated with regions of strong particle acceleration in real separator reconnection events.

Having considered the way in which field line reconnection occurs in the separator configuration 
we proceed next to a quantitative analysis where we ask how to measure
and interpret the rate of reconnection in the configuration with one or more separators.
The question here is whether it is necessary to know the global field topology (including the location and number 
of magnetic separators) in order to determine the reconnection rate.

\section{Measuring Reconnection Rates}
\label{sec:rates}

In a two-dimensional configuration where reconnection takes place at an X-type null point of the field
the reconnection rate is given by the value of the electric field at the null point and measures the rate
at which magnetic flux is transferred between the four topologically distinct flux domains.
In order to express the rate as a dimensionless quantity that electric field is normalised to a characteristic 
convective electric field and so the reconnection rate measured in terms (fractions) of the Alfv{\'e}n Mach number.

In three-dimensions we also have a measure for the rate of reconnection.
This is given by the maximum integrated parallel electric field over all field lines that thread the non-ideal region
(Schindler {\it et al.}~1988):
\begin{equation}
\label{eq:3drec}
 \frac{ \textrm{d} \Phi_{\textrm{rec}}}{ \textrm{d} t} = \left \vert \int E_{\parallel} d l \right \vert_{\textrm{max}}. 
 \end{equation}
 However, the interpretation of the (maximum) integral as a unique reconnection rate
 relies on the assumption that the topology of the magnetic field in the reconnection region is simple.
This assumption is justified in our case only up to the time of the bifurcation of separators.
The formulation \eqref{eq:3drec} is consistent with the two-dimensional measure of the electric field at the null
with the two-dimensional reconnection rate being the three-dimensional reconnected flux 
per unit length in the invariant direction.
Note that the question of how or whether to normalise the three-dimensional reconnection rate \eqref{eq:3drec}
has not been properly addressed.
In contrast to the two-dimensional case a `high' reconnection rate can be obtained by having a strong
value of the electric field or a long non-ideal region (long path of the integral) and the question of what
constitutes `typical'  convective electric field, or field strength to normalise it to, is not easily answered.

To examine the reconnection rate in the model presented here, we first consider only the early stages
of the evolution, $t \in [0,0.440]$, when a single separator is present (along the $x=y=0$ line and with $z \in [-5,5]$).  
During this time the field line with the field line with the maximum integrated parallel electric field along it is the 
separator field line itself.
The linear increase in time of the flux ring (simulating the reconnection) implies that the electric
field at this line is constant in time and so the reconnection rate of the configuration for
$t \in [0,0.440]$ is given by
\[ 
 \frac{ \textrm{d} \Phi_{\textrm{rec, 0}}}{ \textrm{d} t} = \left \vert \int_{z=-5}
 ^{z=5} E_{\parallel}(0,0,z) \ d l \right \vert = 10 \sqrt{\pi} \ \textrm{erf}(5)
  \approx 17.725.
\]
In this case the coincidence of the separator and the line of maximum parallel electric field gives the clear and intuitive 
interpretation of the reconnection rate as the rate at which flux is transferred between the topologically distinct flux 
domains (from regions II and IV and into regions I and III).

\begin{figure}[htbp] 
   \centering
      \includegraphics[width=0.99\textwidth]{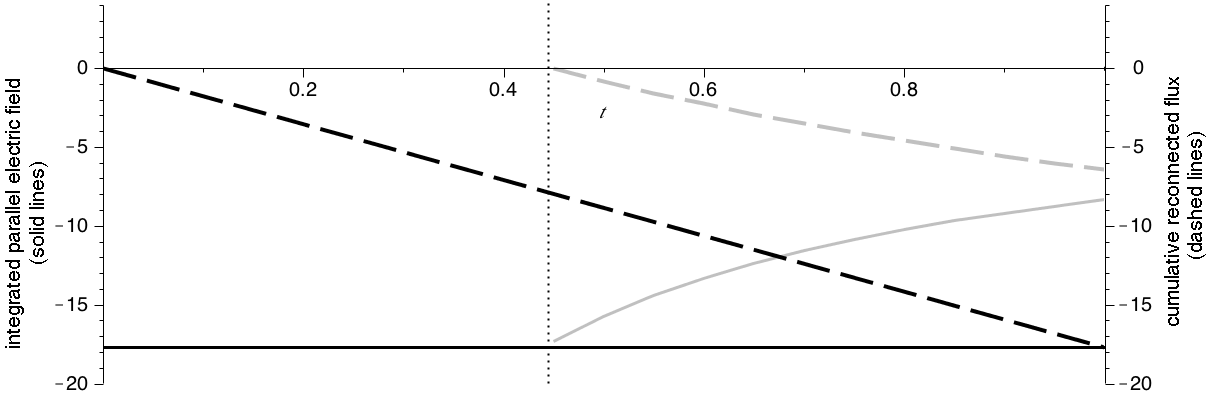} 
   \caption{Integrated parallel electric field across the central separator $S_{0}$ (solid black line) and 
   the separators $S_{1}$, $S_{2}$ (solid grey line) that form in the first bifurcation.  The cumulative reconnected
   flux across the separators is shown in the  dashed lines of corresponding colour. The 
   dotted vertical line indicates the moment of bifurcation of separators.}
   \label{fig:recnratedata}
\end{figure}

Following the first bifurcation three separators are present in the domain (and five following the second bifurcation);
we aim to determine how the reconnection rate should be measured and interpreted in this situation. 
We label the central separator $S_{0}$ and the two separators lying off the central axis following
the first bifurcation $S_{1}$ and $S_{2}$ (by symmetry the order is not important).  The values of
the quantities
\[ \frac{ \textrm{d} \Phi_{\textrm{rec, 0}}}{ \textrm{d} t} = \int_{S_{0}} E_{\parallel} d l
\ \ \ \text{and} \ \ \ 
 \frac{ \textrm{d} \Phi_{\textrm{rec, 1}}}{ \textrm{d} t} =  \int_{S_{1}} E_{\parallel} d l
 =  \int_{S_{2}} E_{\parallel} d l =  \frac{ \textrm{d} \Phi_{\textrm{rec, 2}}}{ \textrm{d} t}\]
over time are shown in Figure~\ref{fig:recnratedata} along with the associated cumulative reconnected fluxes
(time integrals of the reconnection rates).  For $t \in [0.440,1.0]$
 a total of $\phi_{1} = \phi_{2} = 6.453$ units of flux are reconnected through 
each of the separators $S_{1}$ and $S_{2}$.

The physical interpretation for the fluxes $\phi_{1}$ and $\phi_{2}$ comes from considering the difference in 
reconnected flux between the separators $S_{0}$ and $S_{1}$ (or, equivalently, $S_{2}$) in the interval $[0.440,1.0]$.
This is given by
\begin{equation}
\label{eq:inoutinout}
- \int_{t=0.440}^{1}  \frac{ \textrm{d} \Phi_{\textrm{rec, 0}}}{ \textrm{d} t} \ d t \ +  
\int_{t=0.440}^{1} \frac{ \textrm{d} \Phi_{\textrm{rec, 1}}}{ \textrm{d} t} \ d t 
= 0.56 \times 17.725 - 6.453  \approx 3.473. \end{equation}
This value is that of the magnetic flux passing through the surface bounded by the separators $S_{1}$ and $S_{0}$
at the end of the reconnection and so the flux contained in each of the new flux domains 
(regions V and VI) at the end of the reconnection ($t=1$).  

\begin{figure}[htbp] 
   \centering
      \includegraphics[width=0.45\textwidth]{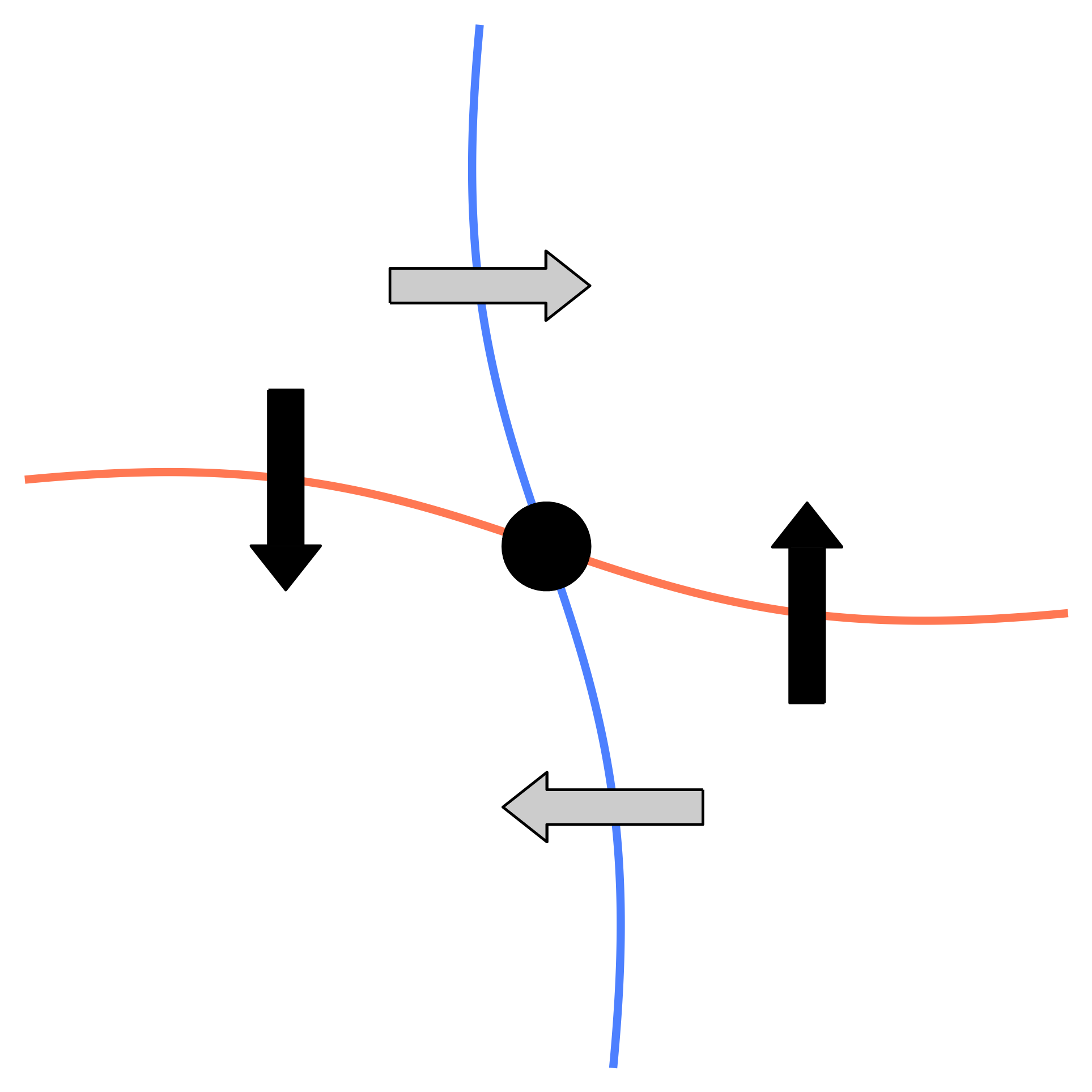} 
            \includegraphics[width=0.45\textwidth]{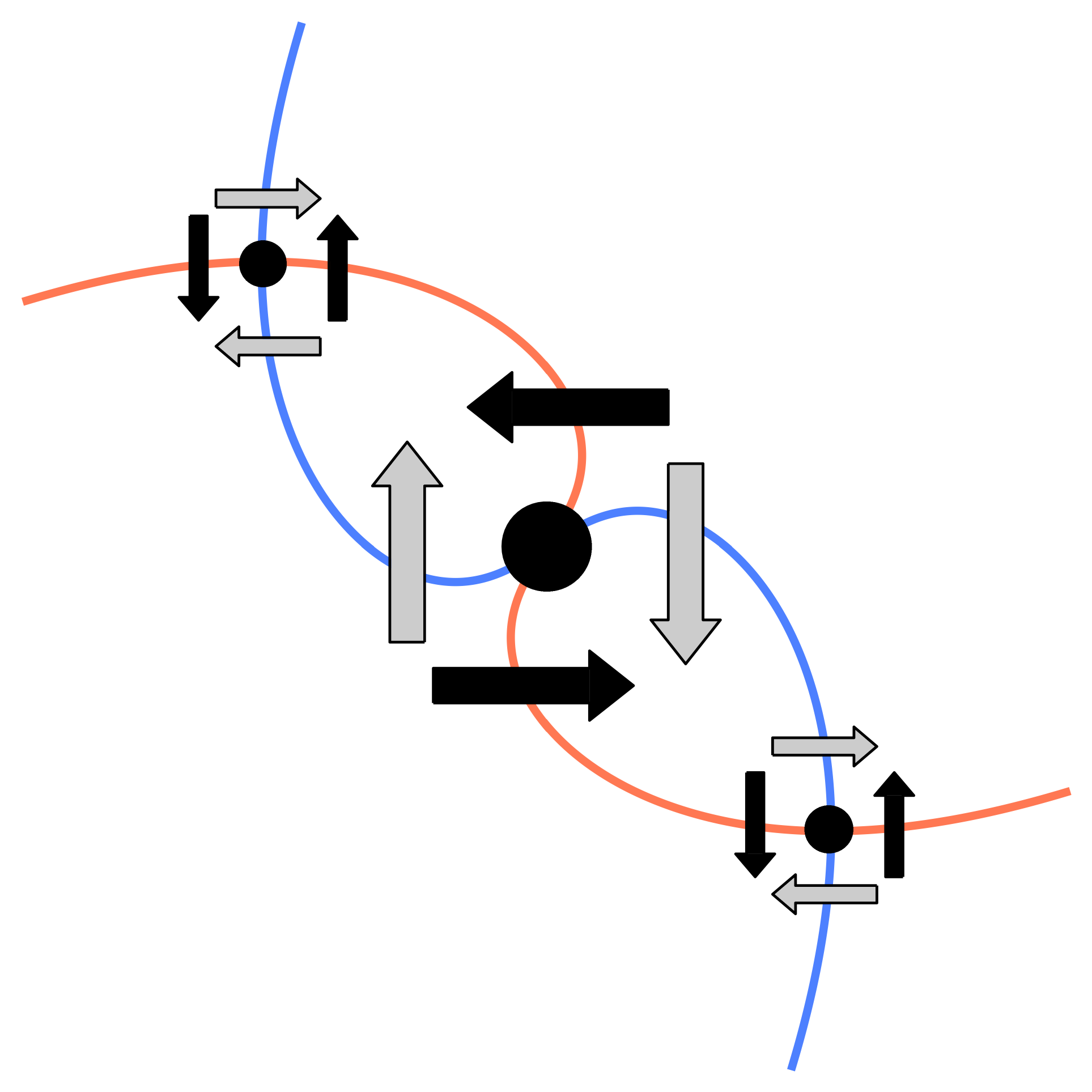} 
   \caption{Directions of flux transport between domains in the case of one separator (left) and 
   multiple separators (right).}
   \label{fig:recnrateinterpret}
\end{figure}

Overall, the situation is illustrated in Figure~\ref{fig:recnrateinterpret} where the flux transport
between domains (across fan surfaces) is shown.
In the figure the black arrows show flux transport 
in the situation where field lines are considered as
 as fixed from below the non-ideal region.  The alternative, the flux transport
 obtained by considering field lines as fixed above the non-ideal region, is shown in grey.
When one separator is present (left-hand image) the integrated parallel electric field along that separator
gives information on flux transport between the four domains in the relatively simple way already discussed.
When multiple separators are present (right-hand image) the separator with the highest integrated parallel electric field
provides the primary transport of flux while the secondary separators provide additional information on the
amount of flux in each domain.  
Considering flux in domain VI for example (the lower closed domain), 
the transport of flux into the domain (thick grey and black arrows) associated with the primary, central separator 
is faster than transport out of the domain (thinner grey and black arrows) associated with the 
secondary bounding separator.  This results in a net transport of flux into this region.

When three separators are present the reconnection rate with respect to the four basic flux 
connectivity types (corresponding to flux domains I, II/VI, III, IV/V, as described in Section~\ref{sec:bif}) is given by
\begin{equation}
\label{eq:threeseprate}
\frac{\textrm{d} \Phi_{\textrm{rec, 1}}}{\textrm{d} t} +\frac{ \textrm{d} \Phi_{\textrm{rec, 2}}}{ \textrm{d} t}  
-\frac{ \textrm{d} \Phi_{\textrm{rec, 0}}}{ \textrm{d} t}. 
\end{equation}
For ${\textrm{d}  \Phi_{\textrm{rec, 0}}}/{ \textrm{d} t} > 
 {\textrm{d}  \Phi_{\textrm{rec, 1}}}/{ \textrm{d} t}, {\textrm{d}  \Phi_{\textrm{rec, 2}}}/{ \textrm{d} t}$,
the situation considered here, this rate is lower than ${\textrm{d}  \Phi_{\textrm{rec, 0}}}/{ \textrm{d} t}$.=
The reduction in the rate of change of flux between domains compared with the reconnection rate
as given by the maximum integrated parallel electric field across the region (${\textrm{d}  \Phi_{\textrm{rec, 0}}}/{ \textrm{d} t}$)
is a result of recursive reconnection between the multiple separators.  
The recursive nature of the reconnection can be seen in Figure~\ref{fig:recnrateinterpret}.  As an example 
considering flux as fixed from below the non-ideal region (i.e.~with the grey arrows),  the amount of flux in 
domain  I is being reduced at a rate ${\textrm{d}  \Phi_{\textrm{rec, 0}}}/{ \textrm{d} t}$ by reconnection into region VI but also 
increased at a rate ${\textrm{d} \Phi_{\textrm{rec, 1}}}/{ \textrm{d} t}$  from domain VI and rate 
${\textrm{d}  \Phi_{\textrm{rec, 2}}}/{ \textrm{d} t}$ 
from domain II.

Note that the situation ${\textrm{d}  \Phi_{\textrm{rec, 0}}}/{ \textrm{d} t} < 
{\textrm{d}  \Phi_{\textrm{rec, 1}}}/{ \textrm{d} t}, {\textrm{d}  \Phi_{\textrm{rec, 2}}}/{ \textrm{d} t}$
is also conceivable.  In this case the flux in the regions $V$ and $VI$ decreases in time
and the additional flux coming from these regions can push the reconnection rate \ref{eq:threeseprate}
above the maximum of $\{ {\textrm{d}  \Phi_{\textrm{rec, 0}}}/{ \textrm{d} t},
{\textrm{d}  \Phi_{\textrm{rec, 1}}}/{ \textrm{d} t}, {\textrm{d}  \Phi_{\textrm{rec, 2}}}/{ \textrm{d} t} \}$.
Both situations show that due to the non-trivial topology of the magnetic field the reconnection rate
with respect to the flux domains can differ significantly from the reconnection rate \ref{eq:3drec}.  This is 
of particular relevance for the measurement and interpretation of reconnection since separators can be
hard to detect in real systems (and numerical simulations).

We remark further that if one is, for example, interested in particle acceleration rather than flux evolution,
it may not be important that the reconnection processes can cancel each other in a topological sense.  In this case
the sum of all the reconnection rates,
\[ \left\vert \frac{{\textrm{d}  \Phi_{\textrm{rec, 0}}}}{{ \textrm{d} t}} \right\vert +
\left \vert  \frac{{\textrm{d}  \Phi_{\textrm{rec, 1}}}}{{ \textrm{d} t}}  \right \vert +
\left \vert  \frac{{\textrm{d}  \Phi_{\textrm{rec, 2}}}}{{ \textrm{d} t}}  \right \vert, \]
would give a better measure for the efficiency of the reconnection process.

\section{Discussion: Breaking the Symmetry}
\label{sec:discussion}

The preceding analysis examined a rather particular situation in which a reconnection process was centred
exactly on a magnetic separator that was identified as the reconnection line.
A natural question arises:  is a real reconnection event likely to be centred in this way or will it perhaps
encompass a separator but with a different magnetic field line being the reconnection line?
It seems likely that both cases will occur in real reconnection events and so we now consider how
our findings are altered by centring the reconnection   away from the initial separator.
Such a modification to the model can easily be made by taking $x_{c}$, $y_{c}$, $z_{c}$ as not all zero
in equation~\eqref{eq:fluxring} and making the appropriate adjustments to equation~\eqref{eq:timeevolution}.

\begin{figure}[thbp] 
\centering
\includegraphics[width=0.42\textwidth]{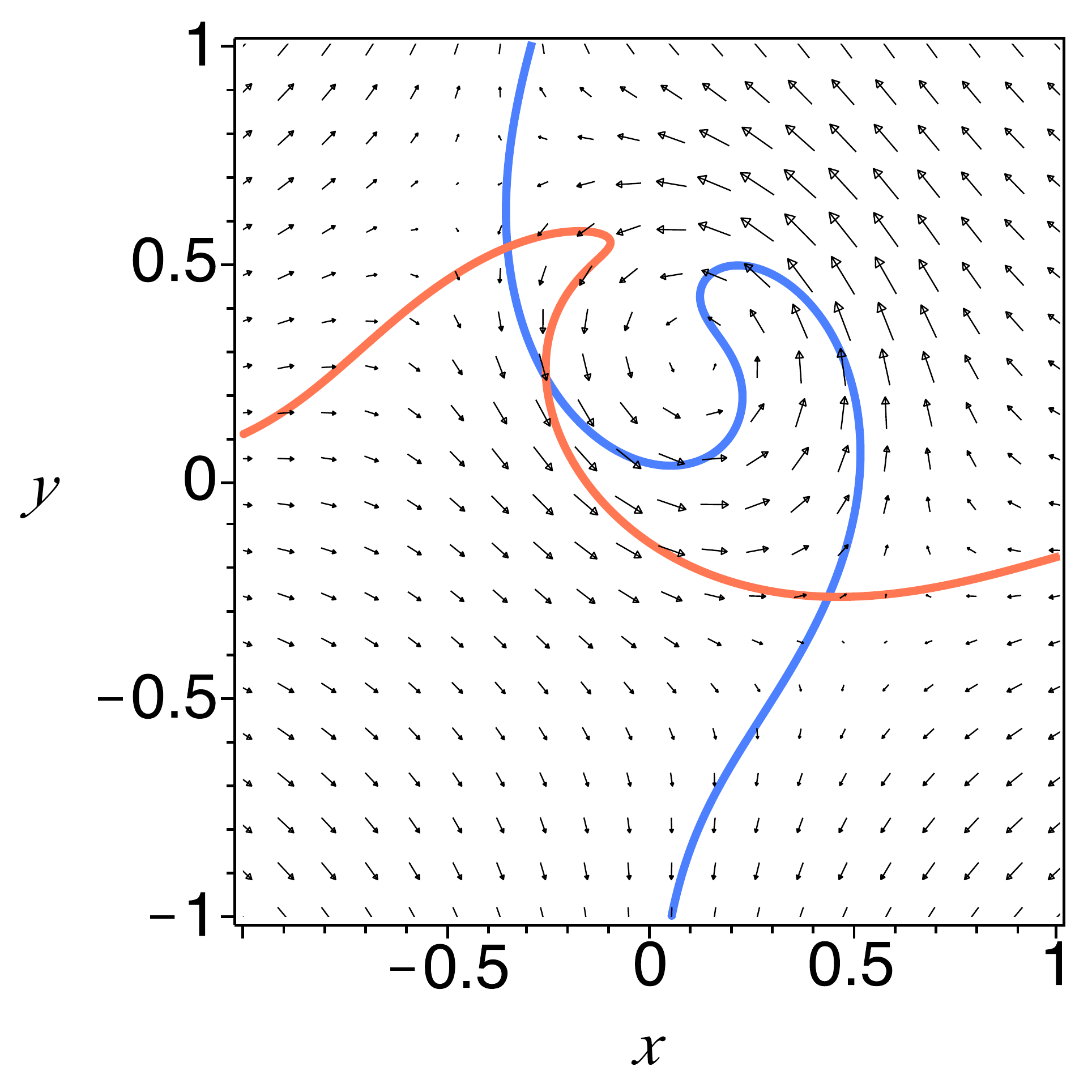} 
\includegraphics[width=0.451\textwidth]{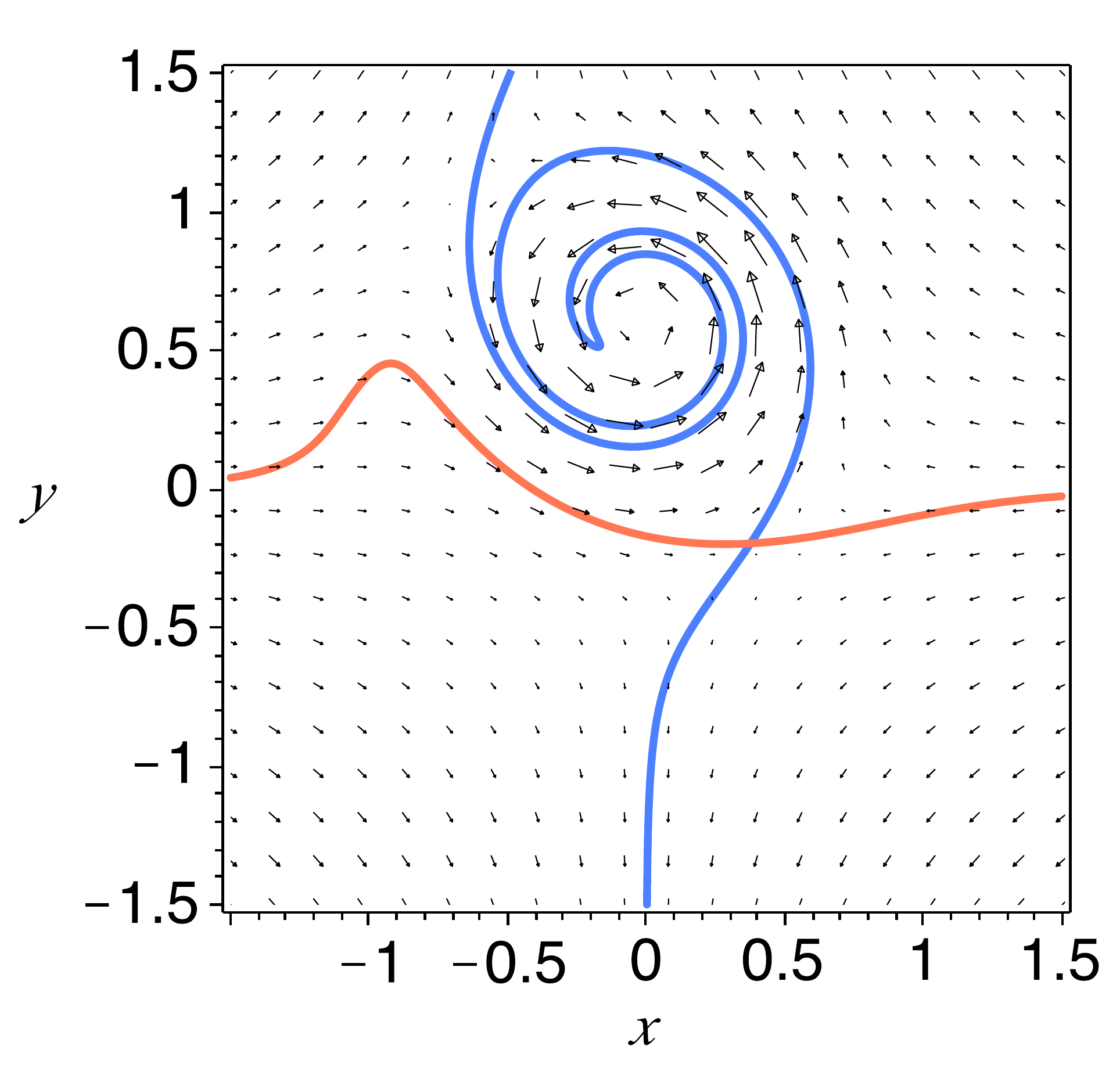} 
\caption{
The topological effect of reconnection at various locations within the basic separator configuration 
using the same method as Figure~\ref{fig:fansz0}.  In the left-hand image the reconnection
region is centred at  $(0.15, 0.3, 0)$ and overlaps the initial separator.  Two more separators are created
as the fan surfaces intersect.  In the right-hand image the reconnection region is centred at $(0, 0.6,0)$
on a fan surface but not including the separator.  No new separator is created in this process.}
   \label{fig:offaxis}
\end{figure}

A first question surrounds our finding that reconnection at a separator can create new separators.
By choosing different locations ($x_{c}, y_{c}$) at which to centre the reconnection region we can further
address the question of whether reconnection in the vicinity of a separator tends to create new
separators. We find that this is indeed the case so long as the reconnection region overlaps the separator
to some degree.  Such an example is shown in Figure~\ref{fig:offaxis} (left).
If the reconnection region lies on or about a fan surface but not including a separator then 
the effect of the reconnection is to twist or wrap up the fan surface (Figure~\ref{fig:offaxis}, right).  
Then only a very particular combination of reconnection events that distort
two separate fan surfaces could lead to an intersection of the surfaces and the creation of a new separator.
We conclude that the creation of new separators by a reconnection event that includes a separator
is a generic process.  The finding may help to explain the large number of separators found in MHD reconnection 
simulations such as those of Parnell {\it et al.} (2010b).

Next we consider the nature of the magnetic flux evolution where a single separator is present but
the separator itself does not give the maximum integrated parallel electric field.
Figure~\ref{fig:sephft2} shows the magnetic flux evolution in the case where the separator and the 
maximum integrated parallel electric field do coincide (the reconnection region is centred at the separator).
Here a rotational flux evolution in the magnetic flux tube threading the non-ideal region splits into 
four `wings'  from the presence of the nulls.
A similar evolution occurs when the reconnection region is offset from the separator.  
In that case the flux tube is split into four wings of unequal sizes but, nevertheless,
the primary features of a continuous change in field line connectivity, a flipping
of magnetic field lines as their topology is changed and the non-local involvement of the null all remain.
That is, it is not necessary for a separator to be the reconnection line in order for the the reconnection 
to have these distinguishing features of separator reconnection.  In this way we can think of separator reconnection 
as a process that takes place when a magnetic separator passes through a reconnection region.
The separator need not itself be associated with the maximum parallel electric field.


\section{Conclusions}
\label{sec:conc}

Separator reconnection is a ubiquitous process in astrophysical plasmas (e.g.~Longcope {\it et al.}~2001;
Haynes {\it et al.}~2007; Dorelli {\it et al.}~2007) but several fundamental details of how it takes place including how flux 
evolves in the process are not yet well understood.
Here we have introduced a simple model for reconnection at an isolated non-ideal region threaded by 
a magnetic separator in an effort to better understand the basics of the process.  Our first new finding is 
that reconnection events in the neighbourhood of a separator can (and if sufficiently strong do) create new separators. 
This occurs even though the null points themselves remain in an ideal region. 
Such a bifurcation of separators has to occur in pairs ({from 1 to} 3,  5, 7, ... separators) and the reverse process
 is, of course, also possible. 
  The fan planes of the two null points divide the space into distinct regions which can be 
 distinguished due to the connectivity of their field lines with respect to any surface enclosing the configuration.  
 Each bifurcation introduces a new pair of flux domains with a connectivity different from their neighbouring fluxes. 

The rate of change of flux between four neighbouring flux domains is usually found 
by the integral of the parallel electric field along their dividing separator. However, in cases where 
multiple separators thread the non-ideal region the change in flux between domains is more complicated, 
as expressed here by Equation~\eqref{eq:threeseprate}, and it will typically be less than the maximum of rates along the 
separators due to recursive reconnection between separators. 
Furthermore, considerations of asymmetric cases show that the maximum of  $\int {\bf E}_\| dl$
does not have to occur along a separator and hence even for a single separator and a single isolated
reconnection region the maximum integrated parallel electric field may not correspond to the rate of change 
of flux between domains.   In  three dimensions the topology of the magnetic field must be known before 
a meaningful reconnection rate can be determined.

Finally, the nature of the magnetic flux evolution in a separator reconnection process shows several distinguishing 
characteristics  including  rotational flux velocities that are split along the separator.
These characteristics are present regardless of whether the separator itself possesses the maximum integrated
parallel electric field.

This analytical model is necessarily limited, one such limitation being in not addressing the way in which 
reconnection could be initiated at a separator.  
Simulations do show current layers forming in the neighbourhood of separators (e.g. Parnell {\it et al.}~2010b)
but the extent of the layers may be variable.
For example, we have assumed here that the non-ideal region is localised to the central portion of the separator 
and the nulls remain in an ideal environment.  If a non-ideal region were to extend further along the domain and 
to include the nulls then 
further bifurcation processes can occur due to bifurcations of null points and a considerably more complex topology can evolve.

\

\noindent
{\bf Acknowledgements}

\noindent
The authors would like to thank an anonymous referee for pointing out the possibility of having a 
reconnection rate with respect to the flux domains which is greater than any of the rates along separators.

\newpage

\begin{center}
{\bf REFERENCES}
\end{center}

\noindent
Albright, B.J. 1999, Phys. Plasmas 6(11), 4222.

\noindent
Birn, J.,  Priest, E.R. 2007,  Reconnection of Magnetic Fields:  Magnetohydrodynamics and Collisionless Theory 
and Observations, Cambridge University Press. 

\noindent
Browning, P.K. Gerrard, C., Hood, A.W., Kevis, R., Van der Linden, R.A.M. 2008,  Astron. Astrophys. 485, 837.

\noindent
Chance, M.S., Greene, J.M., Jensen, T.H. 1992,  Geophys. Astrophys. Fluid Dyn. 64 (1 \& 4), 203.

\noindent
Craig, I.J.D., Fabling, R.B., Heerikhuisen, J., Watson, P.G. 1999, Astrophys. J. 523 (2), 838.

\noindent
Dorelli, J.C., Bhattacharjee, A., Raeder, J. 2007, J. Geophys. Res., 112, A02202.

\noindent
Galsgaard, K.,  Nordlund, A. 1997, J. Geophys. Res., 102, 231.

\noindent
Galsgaard, K., Pontin, D.I. 2011, Astron. Astrophys. 529 A20 .

\noindent
Greene, J.M. 2002,  J. Computational Phys. 98, 194.

\noindent
Haynes, A.L., Parnell, C. E., Galsgaard, K., Priest, E. R. 2007,
Proc. Roy. Soc. A, 463 (2080) 1097.

\noindent
Hesse, M., Schindler, K. 1988, J. Geophys. Res. 93, 5559.

\noindent
Hornig, G., Priest, E.R. 2003, Phys. Plasmas 10(7) 2712.

\noindent
Hornig, G. 2007, `Fundamental Concepts'  in `Reconnection of Magnetic Fields:  Magnetohydrodynamics and Collisionless Theory 
and Observations', eds. Birn, J. and Priest, E.R., Cambridge University Press.

\noindent
Klapper, I., Rado, A., Tabor, M. 1996,  Phys. Plasmas 3, 4281.

\noindent
Lau, Y-T., Finn, J.M. 1990,  Astrophys. J., 350, 672.

\noindent
Longcope, D.W. 1996,   Solar Phys. 169, 91.

\noindent
Longcope, D.W. 2001, Phys. Plasmas 8(12) 5277.

\noindent
Longcope, D.W., Kankelborg, C.C., Nelson, J.L., Pevtsov, A.A. 2001,  Astrophys. J., 553, 429. .

\noindent
Longcope, D. 2007, 
`Separator Reconnection' in `Reconnection of Magnetic Fields:  Magnetohydrodynamics and Collisionless Theory 
and Observations', eds. Birn, J. and Priest, E.R., Cambridge University Press.

\noindent
Galsgaard, K., Parnell, C.E. 2005, Astron. Astrophys. 439, 335.

\noindent
Parnell, C. E., Haynes, A.L., Galsgaard, K. 2008, 
Astrophys. J., 675, 1656.

\noindent
Parnell, C.E., Haynes, A.L.,
Galsgaard, K. 2010a,  J. Geophys. Res., 115, A02102.

\noindent
Parnell, C. E.; Maclean, R. C.; Haynes, A. L. 2010b, Astrophys. J. Letters, 725(2) L214.

\noindent
Pontin, D.I., Hornig, G., Priest, E.R. 2004, 
Geophys. Astrophys. Fluid Dyn. 98, 407.

\noindent
Pontin, D.I., Hornig, G., Priest, E.R. 2005, 
Geophys. Astrophys. Fluid Dyn. 99, 77.

\noindent
Priest, E.R., Pontin, D.I. 2009,  Physics of Plasmas 16 122101..

\noindent
Priest, E.R., Titov, V.S. 1996,  Phil. Trans. Roy. Soc. A,
354 (1721) 2951.

\noindent
Schindler, K., Hesse, M., Birn, J. 1988,  J. Geophys. Res.
93, 5547.

\noindent
Titov, V.S., Hornig, G., D{\'e}moulin, P. 2002,
J. Geophys. Res. 107(8).

\noindent
Wilmot-Smith, A.L., Hornig, G., Priest, E.R. 2006,  Proc. Roy. Soc. A, 462, 2877.

\noindent
Wilmot-Smith, A.L., Hornig, G., Priest, E.R. 2009, Geophys. Astrophys. Fluid. Dyn.  103 (6) 515.

\noindent
Wilmot-Smith, A.L., Pontin, D.I., Hornig, G. 2010, 
Astron. Astrophys. 516 A5.

\noindent
Xiao, C.J. and 17 others 2007, 
Nature Phys. 3, 609.

\newpage

\end{document}